%% file: IRS_UAV_flexible_hUAV_v5_clean.tex
\documentclass{ieeeaccess}
\usepackage{cite}
\usepackage{amsmath,amssymb,amsfonts}
\usepackage{algorithmic}
\usepackage{graphicx}
\usepackage{textcomp}

\usepackage{multirow}
\usepackage{booktabs}
\usepackage{amsfonts}
\usepackage{placeins}
\usepackage{array}
\usepackage{amsmath,cases,amssymb}
\usepackage{graphicx}
\usepackage{cite}
\usepackage{subfigure}
\usepackage{epsfig}
\usepackage{algorithm}
\usepackage{epstopdf}
\usepackage{balance}
\usepackage{color,soul}

\newcommand{\subparagraph}{}

\newtheorem{mylem}{\bf Lemma}

\newcommand{\subjectto}{{\mathrm{s.t.}}}

\newcommand{\calA}{{\mathcal {A}}}

\newcommand{\calK}{{\mathcal {K}}}
\newcommand{\calM}{{\mathcal {M}}}
\newcommand{\calU}{{\mathcal {U}}}
\newcommand{\non}{\nonumber}
\newcommand{\subnum}{\IEEEyessubnumber}
\newcommand{\bw}{\mathbf {w}}
\newcommand{\bd}{\mathbf {d}}

\newcommand{\bg}{\mathbf {g}}
\newcommand{\barg}{\bar{\mathbf {g}}}
\newcommand{\bG}{\mathbf {G}}
\newcommand{\barG}{\bar{\mathbf {G}}}

\newcommand{\bbeta}{\boldsymbol{\beta}}
\newcommand{\bmu}{\boldsymbol{\mu}}
\newcommand{\btau}{\boldsymbol{\tau}}
\newcommand{\brho}{\boldsymbol{\rho}}

\newcommand{\bzeta}{\boldsymbol{\zeta}}

\newcommand{\BS}{\mathtt{BS}}
\newcommand{\IRS}{\mathtt{IRS}}
\newcommand{\UAV}{\mathtt{UAV}}
\newcommand{\UEk}{\mathtt{UE}_{k}}
\newcommand{\UE}{\mathtt{UE}}
\newcommand{\UEell}{\mathtt{UE}_{\ell}}

\newcommand{\bomega}{\boldsymbol{\omega}}

\newcommand{\bLambda}{\boldsymbol{\Lambda}}
\newcommand{\bPhi}{\boldsymbol{\Phi}}
\newcommand{\bdelta}{\boldsymbol{\delta}}
\newcommand{\btheta}{\boldsymbol{\theta}}
\newcommand{\blambda}{\boldsymbol{\lambda}}
\newcommand{\bvarphi}{\boldsymbol{\varphi}}

\newcommand{\Pmax}{P_{\mathtt{BS}}^{\max}}

\newcommand{\fmul}{f_{\mathtt{mul}}}
\newcommand{\fpow}{f_{\mathtt{pow}}}
\newcommand{\fqu}{f_{\mathtt{qu}}}
\newcommand{\fsk}{f_{\mathtt{s},k}}
\newcommand{\fnk}{f_{\mathtt{n},k}}

\newcommand{\Fsr}{F_{\mathtt{SR}}}
\newcommand{\Fsrbar}{\bar{F}_{\mathtt{SR}}}
\newcommand{\FPL}{F_{\mathtt{PL}}}
\newcommand{\FRa}{F_{\mathtt{Ra}}}

\newcommand{\alphaBSUAV}{\alpha_{\mathtt{BS} ,\mathtt{UAV} }}
\newcommand{\alphaUAVUE}{\alpha_{\mathtt{UAV} , \mathtt{UE}} }

\newcommand{\buUAV}{\mathbf{u}_{\mathtt{UAV}}}

\newcommand{\Upper}{\mathtt{up}}
\newcommand{\Lower}{\mathtt{low}}

\newcommand{\hUAV}{h_\mathtt{UAV}}

\newcommand{\hBS}{h_\mathtt{BS}}
\newcommand{\rUAVBS}{r_\mathtt{UAV,BS}}

\def\BibTeX{{\rm B\kern-.05em{\sc i\kern-.025em b}\kern-.08em
    T\kern-.1667em\lower.7ex\hbox{E}\kern-.125emX}}
\begin{document}
\doi{10.1109/ACCESS.2017.DOI}

\title{Joint UAV Placement and IRS Phase Shift Optimization in Downlink Networks}
\author{\uppercase{Hung Nguyen-Kha}\authorrefmark{1}, \uppercase{Hieu V. Nguyen}\authorrefmark{2}, \uppercase{Mai T. P. Le}\authorrefmark{2}, and \uppercase{Oh-Soon Shin}\authorrefmark{1}}
\address[1]{School of Electronic Engineering \& Department of ICMC Convergence Technology, Soongsil University, Seoul 06978, Korea}
\address[2]{Faculty of Electronics and Telecommunication Engineering, The University of Danang - University of Science and Technology, Da Nang, Viet Nam}
\tfootnote{This paragraph of the first footnote will contain support 
information, including sponsor and financial support acknowledgment. The work of Mai T. P. Le was funded by Vingroup JSC and supported by the Postdoctoral Scholarship Programme of Vingroup Innovation Foundation (VINIF), Institute of Big Data, code VINIF.2021.STS.19.}

\markboth
{Author \headeretal: Preparation of Papers for IEEE TRANSACTIONS and JOURNALS}
{Author \headeretal: Preparation of Papers for IEEE TRANSACTIONS and JOURNALS}

\corresp{Corresponding author: Oh-Soon Shin (e-mail: osshin@ssu.ac.kr).}

\begin{abstract}
This study investigates the integration of an intelligent reflecting surface (IRS) into an unmanned aerial vehicle (UAV) platform to utilize the advantages of these leading technologies for sixth-generation communications, e.g., improved spectral and energy efficiency, extended network coverage, and flexible deployment. In particular, we investigate a downlink IRS--UAV system, wherein single-antenna ground users (UEs) are served by a multi-antenna base station (BS). To assist the communication between UEs and the BS, an IRS mounted on a UAV is deployed, in which the direct links are obstructed owing to the complex urban channel characteristics. The beamforming at the BS, phase shift at the IRS, and the 3D placement of the UAV are jointly optimized to maximize the sum rate. Because the optimization variables, particularly the beamforming and IRS phase shift, are highly coupled with each other, the optimization problem is naturally non-convex. To effectively solve the formulated problem, we propose an iterative algorithm that employs block coordinate descent and inner approximation methods. Numerical results demonstrate the effectiveness of our proposed approach for a UAV-mounted IRS system on the sum rate performance over the state-of-the-art technology using the terrestrial counterpart.
\end{abstract}

\begin{keywords}
	Beamforming, intelligent reflecting surface (IRS), unmanned aerial vehicle (UAV), UAV-mounted IRS, convex optimization.
\end{keywords}

\titlepgskip=-15pt

\maketitle

\input{Sec_Intro.tex}

\input{Sec_Model.tex}

\input{Sec_Proposed_SR.tex}

\input{Sec_Results.tex}

\input{Sec_Conclusion.tex}


\input{Sec_Appendix.tex}

\bibliographystyle{IEEEtran}
\balance
\bibliography{Journal}

\newpage

\begin{IEEEbiography}[{\includegraphics[width=1in,height=1.25in,clip,keepaspectratio]{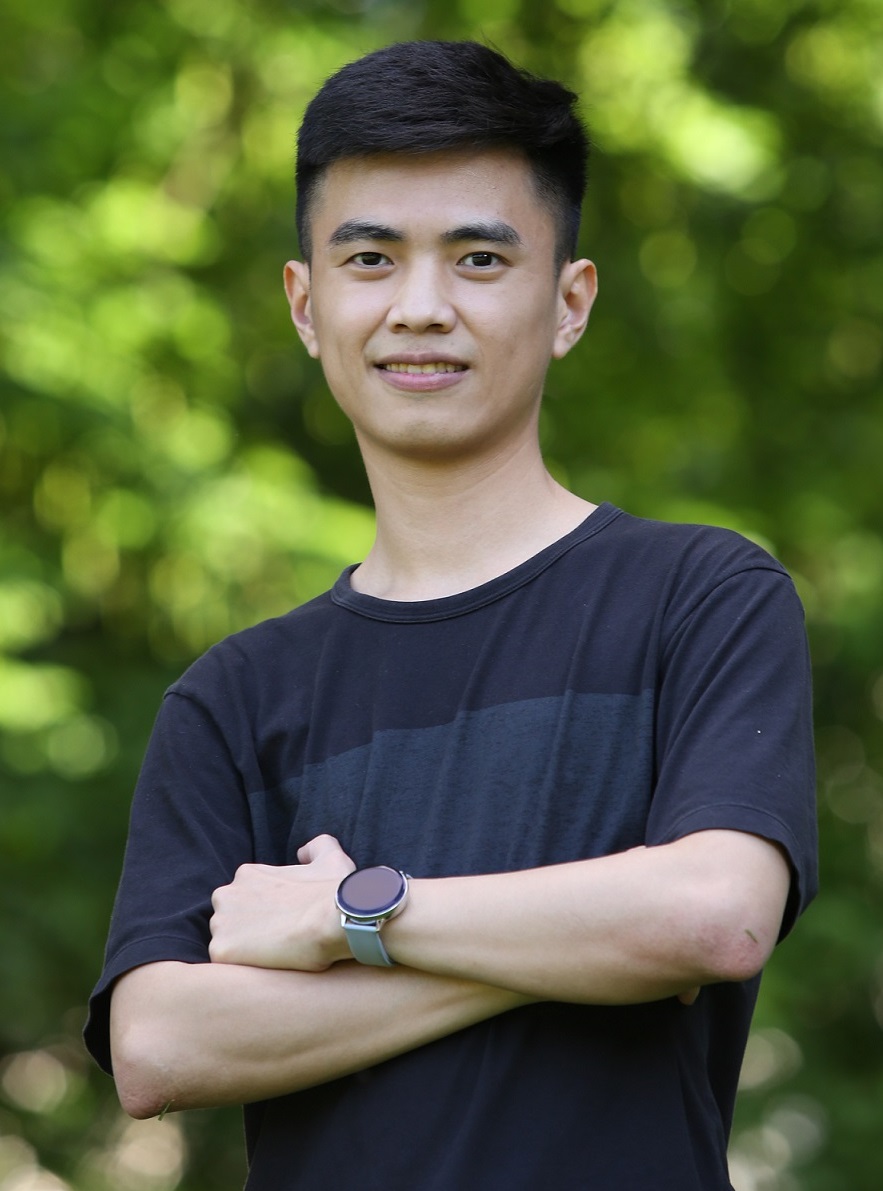}}]{Hung Nguyen-Kha} 
	received the B.E. degree in electronic and telecommunication from Posts and Telecommunications Institute of Technology, Hanoi, Vietnam, in 2019, and the M.S. degree in electrical engineering from Soongsil University, South Korea, in 2021.
	
	From 2019 to 2021, he was with the Wireless Communication Laboratory at Soongsil University. His research interest is in wireless communications, with particular focus on optimization techniques, 5G system, beamforming design, NOMA, and IRS communications.
\end{IEEEbiography}

\begin{IEEEbiography}[{\includegraphics[width=1in,height=1.25in,clip,keepaspectratio]{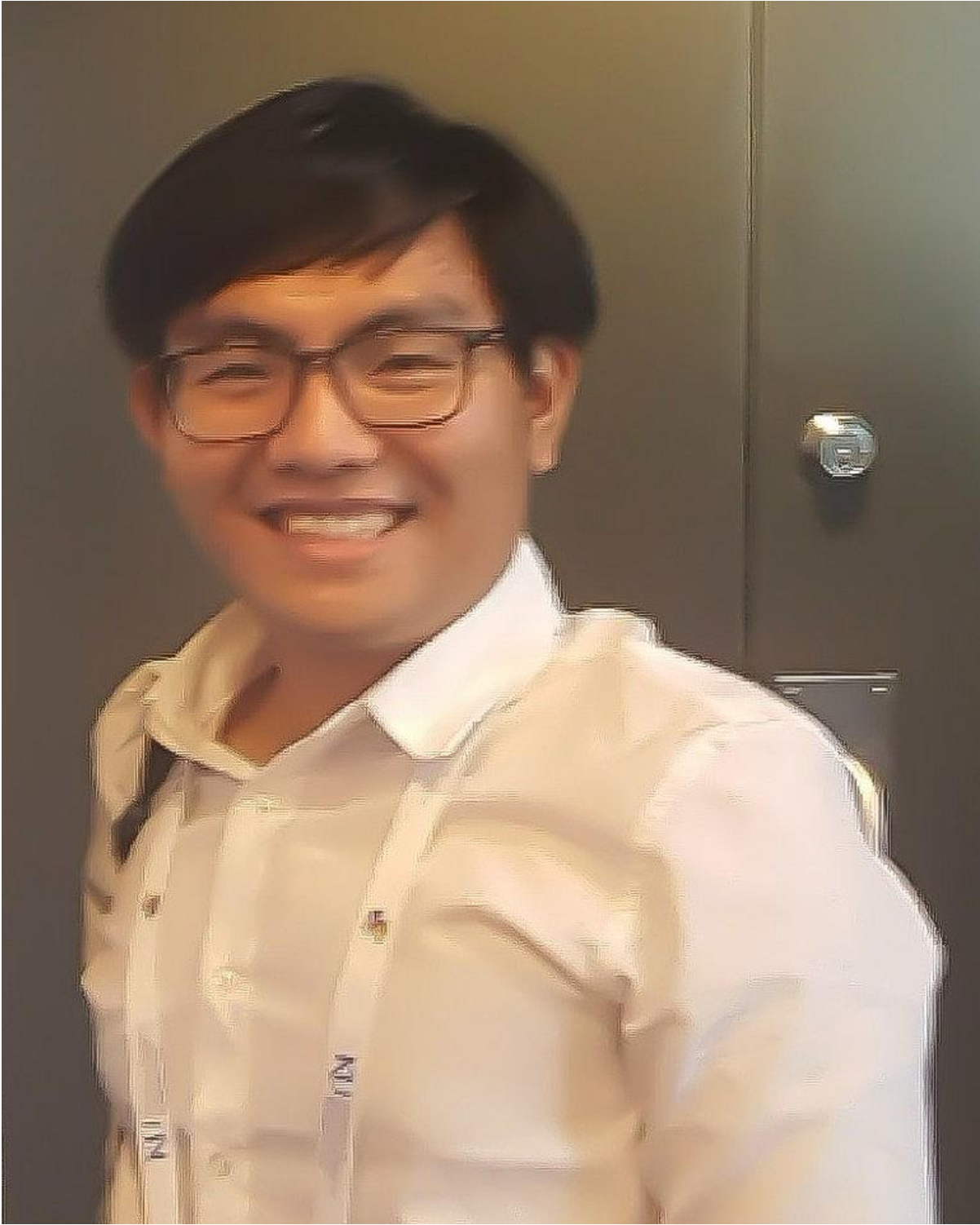}}]{HIEU V. NGUYEN} (S'16-M'21) received the B.E. degree in electronics and telecommunications from The University of Danang - University of Science and Technology (UD-DUT), Vietnam, in 2011, and the M.E. and Ph.D. degrees in electronic engineering from Soongsil University, Seoul, Korea, in 2016 and 2020, respectively. From 2011, he has been with UD-DUT where he is currently a Lecturer.
	
From 2014 to 2021, he was with Wireless Communications Laboratory at Soongsil University as a Research Assistant and a Postdoc Researcher. From 2021 to 2022, he was a Research Associate with the SnT, Interdisciplinary Centre for Security, Reliability and Trust, University of Luxembourg. His research interest is in wireless communications, with particular focus on optimization techniques and machine learning for wireless communications, such as UAV/drones communications, device-to-device communications, full-duplex radios, green communication systems, IoT and satellite networks.
	
\end{IEEEbiography}

\vspace{-20pt}

\begin{IEEEbiography}[{\includegraphics[width=1in,height=1.25in,clip,keepaspectratio]{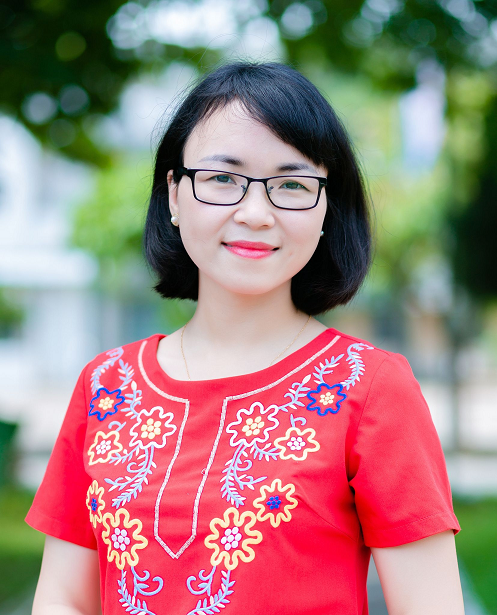}}]{Mai T. P. Le} received the Ph.D. degree from Sapienza University of Rome, Rome, Italy, in February 2019. Since 2011, she has been with the Department of Electronics and Telecommunications, the University of Danang - University of Science and Technology, Da Nang, Vietnam, where she is currently a Lecturer. 
	
From 2015 to 2020, she was a Ph.D. student and Postdoctoral Researcher with the Department of Information Engineering, Electronics and Telecommunications, Sapienza University of Rome. In 2016, she was a Visiting Researcher with the Singapore University of Technology and Design, Singapore in 2016, and in 2012, with the Arizona State University, Tempe, AZ, USA. Her main research interests include information theory, mathematical theories, and their application in wireless communications. Her current research focuses on physical layer techniques for the next generation network.
\end{IEEEbiography}

\begin{IEEEbiography}[{\includegraphics[width=1in,height=1.25in,clip,keepaspectratio]{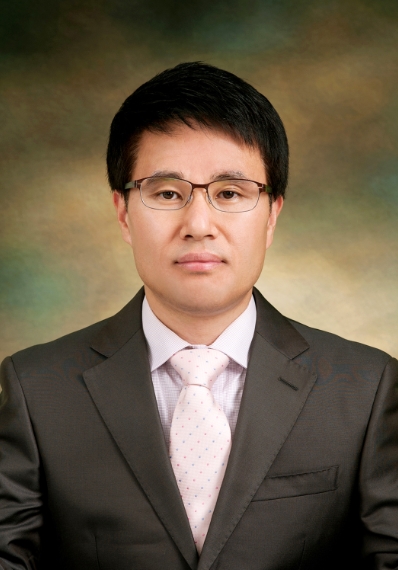}}]{OH-SOON SHIN} (S'00-M'10) received the B.S., M.S., and Ph.D. degrees in electrical engineering and computer science from Seoul National University, Seoul, South Korea, in 1998, 2000, and 2004, respectively. 
	
From 2004 to 2005, he was with the Division of Engineering and Applied Sciences, Harvard University, MA, USA, as a Post-Doctoral Fellow. From 2006 to 2007, he was a Senior Engineer with Samsung Electronics, Suwon, South Korea. In 2007, he joined the School of Electronic Engineering, Soongsil University, Seoul, where he is currently a Professor. His research interests include communication theory, wireless communication systems, and signal processing for communication.
\end{IEEEbiography}

\EOD

\end{document}

%% file: Sec_Intro.tex
\section{Introduction}
Among the disruptive technologies proposed for sixth-generation communications, the intelligent reflecting surface (IRS), also known as a reconfigurable intelligent surface or software-controlled metasurface, has recently emerged as a prominent candidate owing to its ability to completely control the wireless medium between transceivers \cite{Holographic_Surface_6G}. In particular, an IRS is a surface composed of a significant number of passive elements, each of which functions as a smart scatterer, i.e., it reflects the impinging signals and adjusts the phase shifts appropriately via a programmable controller \cite{bjorn19}. Therefore, the signals reflected by each IRS element can be combined with those from other paths to increase the signal strength intended for receivers and mitigate the inter-interference, thereby affording significantly improved spectral efficiency \cite{IRS_aided_wireless_networkJ_reconfigurable_environment}. 
In contrast to the conventional active relay, e.g., amplify-and-forward and decode-and-forward, the IRS is constructed using passive elements, i.e., without requiring RF chains \cite{IRS_aided_wireless_networkJ_reconfigurable_environment}. Furthermore, the physical principle of IRS is based on reflecting impinging signals that do not require power consumption, which makes IRS consume low energy \mbox{\cite{RISvsRelaying}}. With the recent advancement in lumped elements such as micro-electro-mechanical-systems, varactors and PIN diodes, and tunable materials, the IRS made out of metasurfaces can be implemented with low cost, compact size and light weight \cite{hum2013}. All the above-mentioned features render the IRS not only a promising solution for green networks but also a favorable associate for other technologies.

Meanwhile, unmanned aerial vehicles (UAVs) have been widely adopted in the last decade for multiple applications, such as surveillance, tracking, remote sensing, and particularly disaster communications (e.g., search and rescue missions after earthquakes/floods), owing to their flexible deployment, autonomy, and cost efficiency \cite{Zeng19}. 
Owing to the high altitude that provides the line-of-sight link, UAVs are favorable for cellular networks, where they can be employed as aerial relays or flying BSs to provide coverage extension to the existing wireless infrastructure as well as connectivity to areas with obstructed direct links or local traffic congestion \cite{Huo19}.

However, the use of UAVs in high-speed transmission networks remains limited owing to their size, weight, and power supply \cite{Zeng19}. With significant improvements in terms of both spectral efficiency (SE) and energy efficiency (EE), the IRS is expected to compensate the abovementioned weakness of UAVs. In particular, by reflecting the impinging signals, the IRS consumes much less power than the conventional signal processing for a BS. This will improve the energy efficiency of a UAV-based IRS system by letting the UAV transmitter in sleep mode if the quality of service (QoS) constraints are satisfied by the IRS-sole mode \cite{Mozaffari19}. Owing to the easy installation and conformal geometry, IRSs can be mounted on the facade of buildings or directly on the UAVs to support terrestrial UEs with limited connectivity. The integration of IRS and UAVs, therefore, has received substantial interest from the community \cite{Li_WCL_20, Wei_TWC_21,MU_JSAC_21,Ge_Access_20, UAV_IRS_Symbiotic_System, PangTCOM21, Sun_TVT_21, Pan_WCL_21, Shafique_TCOM_21,Mahmoud_TGN_21,PangWCL21,Zhou21,Lu_TWC_21}. 

\subsection{Related Work}
In \cite{Li_WCL_20}, a downlink (DL) IRS-assisted UAV system was investigated for a single antenna ground user (UE) scenario, where the phase shift at the IRS and the UAV trajectory were alternately optimized to enhance the SE. The maximized sum rate of a multiuser IRS-aided UAV system, supported by orthogonal frequency division multiple access (OFDMA), was investigated in \cite{Wei_TWC_21}, incorporating with the constraint of the UE QoS. By contrast, the authors of \cite{MU_JSAC_21} employ non-orthogonal multiple access (NOMA) to serve multiple groups of UEs in an IRS-assisted multi-UAV network. In that study, each group of UEs is served by a common IRS and a separate UAV, and the decoding order of each UE group is based on the UE channel gains. To enhance the system throughput, the UAV position, power allocation at each UAV, and phase shift at the IRS are optimized alternately. Another IRS--UAV model employing a UAV and several IRSs to serve a ground UE was investigated in \cite{Ge_Access_20, UAV_IRS_Symbiotic_System}. The maximization of the received power at the UE was used as the target objective in \cite{Ge_Access_20}, whereas the aim of \cite{UAV_IRS_Symbiotic_System} was to achieve the minimum weighted sum bit error rate among the IRSs. For secure communications, the transmit beamforming, the UAV trajectory and the IRS passive beamforming have been jointly considered to maximize the secrecy rate of an IRS-assisted UAV system in \cite{PangTCOM21}. For millimeter-wave-based and THz-based networks, the performance of an integrated IRS--UAV systems was analyzed in \cite{Sun_TVT_21} and \cite{Pan_WCL_21}, respectively.

It is noteworthy that the aforementioned studies share the same feature of an unconnected IRS--UAV model, i.e., the UAV functions as an aerial BS, whereas the IRS is typically placed distant from the UAV, e.g., mounted on the facade of buildings or placed in the surrounding area of a UAV/UE. Based on this paradigm, there have been a number of recent works for the UAV-mounted IRS, where the IRS is mounted on/carried by the UAV and functions as a smart reflector \cite{Shafique_TCOM_21,Mahmoud_TGN_21,PangWCL21,Zhou21,Lu_TWC_21}. Such an IRS-attached UAV model can enjoy the full three-dimensional (3D) reflection while freely traveling in space to satisfy terrestrial communication demands. This may not substantially increase the network coverage, particularly at the cell edges and dead zones, but enables a much lower number of cellular BSs, thereby realizing a green network. 
Notice that a majority of those works, e.g., \cite{Shafique_TCOM_21,Mahmoud_TGN_21,PangWCL21,Zhou21}, focus on supporting the transmission between a single UE and a BS. Specifically, the authors of \cite{Shafique_TCOM_21} derived analytic derivations for the outage probability, SE, and EE, as well as formulated the EE optimization for a system with a UAV carrying an IRS to support point-to-point communication between a BS and a single ground UE. In fact, different transmission modes of the UAV and/or IRS were investigated sequentially. Theoretical analyses pertaining to the symbol error rate, ergodic capacity, and outage probability were performed for a UAV-attached IRS system to support the wireless transmission between a BS and a single UE in \cite{Mahmoud_TGN_21}. On the other hand, the work \cite{PangWCL21} aimed to improve the achievable rate of a target UE by optimizing the trajectory of the UAV-mounted IRS with the given locations of the UE and BS.
The achievable user rate is also the objective of the formulated optimization problem for an aerial IRS-aided cell-free massive MIMO system in \cite{Zhou21}, where the IRS is used to support the transmission between a single UE with limited connection (UE in shadow area) and the access points (APs). Meanwhile, the authors of \cite{Lu_TWC_21} proposed employing an aerial IRS (an IRS mounted on an aerial platform such as a balloon or UAV) to extend the signal coverage from a ground BS to a specific area, targeting the maximization of the worst-case signal-to-noise-ratio in the considered area.

\subsection{Motivation and Main Contributions}{
In this work, we propose a DL IRS-aided UAV network, in which an IRS is designed to be mounted on the UAV for reuse in multiple hot-spot areas without constructing a new infrastructure. Note that the size of one IRS element is typically of range $\lambda/10-\lambda/5$, where $\lambda$ is the wavelength of the transmit wave \cite{liaskos2018}. Therefore, it is feasible for a UAV to carry a moderate-size IRS with an adequate number of elements. Owing to the conformal geometry, IRS provides a potential solution for aerial platforms, as compared to the conventional relays using active components. Such an aerial IRS-UAV model is particularly useful in several scenarios involving overloaded communication traffic, e.g., crowded areas, stations, stadiums or grand events, wherein flexibility and mobility are important factors of the system deployment. 

In this work, we consider a practical scenario where a multi-antenna BS serves multiple UEs via an aerial IRS-UAV relay. While the achievable rate of a single UE in \cite{PangWCL21,Zhou21} or the minimum SNR over all locations within a target area in \cite{Lu_TWC_21} are the system performance objectives, we instead aim at maximizing the average system sum rate, which is naturally a crucial system metric for 6G \cite{Holographic_Surface_6G}. More specifically, a joint optimization problem is presented incorporating the BS beamforming, IRS phase shift, and UAV placement while considering the rate threshold constraint. Note that the UAV placement task in this work is to determine both the optimized topographical position and the optimized altitude of the UAV, i.e., the optimized 3D position of UAV. Meanwhile, we employ the digital beamforming, which allows to allocate different powers and phases to different antennas,  to further enhance degrees of freedom (DoFs) in precoding \cite{survey2022mmWave}. Owing to the coupling of the optimization variables, particularly the beamforming and IRS phase shift, the sum rate problem is nonconvex, which is non-trivial to solve directly. To effectively address the formulated problem, we divide the original problem into two sub-problems, wherein the beamforming and IRS phase shift are alternately optimized with the UAV positions in each sub-problem. Accordingly, we propose an iterative algorithm using the inner approximation (IA) and block coordinate descent (BCD) methods to alternately solve these sub-problems until reaching the convergence.

The main contributions of this study are summarized as follows:
\begin{itemize} 
	\item We investigate a DL MU-MISO network, where UEs are located within a blocked region, i.e., without direct links between the BS and UEs, due to the complex urban environment. Taking advantage of both IRS and UAV technologies, we propose to deploy an IRS mounted on a UAV to assist the terrestrial communications, wherein a general system framework for multi-antenna BS serving multiple UEs with the assistance of a UAV-mounted IRS is presented.
	\item Aiming at maximizing the system sum rate, we formulate an optimization problem incorporating the UAV positioning with optimized altitude, BS beamforming, and IRS phase shift coefficients, which belongs to a difficult class of nonconvex optimization problem. To address it, we utilize BCD to decompose it into two sub-problems. For the first sub-problem, UAV placement and BS beamforming are jointly considered, whereas UAV placement and IRS phase shift are considered for optimization in the second sub-problem. Towards appealing applications, a low-complexity iterative algorithm based on the IA method is developed to address the nonconvex constraints and objective, which are guaranteed to converge to at least a locally optimal solution.
	\item Finally, extensive simulations are executed to demonstrate the effectiveness of our proposed approach with a UAV-mounted IRS on the sum rate performance  compared with conventional terrestrial systems.
\end{itemize} 

\subsection{Outline and Notations}
The remainder of this paper is organized as follows: Section II describes the system model and the problem formulation. In Section III, an iterative algorithm for solving the sum rate maximization problem is proposed. Sections IV and V present the numerical results, and conclusions, respectively.

\textit{Notation:} Scalars, vectors, and matrices are denoted by italic letters, lowercase letters and uppercase bold letters, respectively. $ \mathbf{x}^{H}$ and $ \|\mathbf{x}\|$ are the Hermitian transpose and Euclidean norm of a complex vector $ \mathbf{x} $, respectively; $ \mathtt{diag}(\mathbf{x}) $ denotes the diagonal matrix created from $ \mathbf{x} $. For the scalar $ x $ and matrix $ \mathbf{X} $, $ |x | $ and $ | \mathbf{X} | $ represent the absolute value of $ x $ and the determinant of $ \mathbf{X} $. $\mathbb{R}$ and $\mathbb{C}$ represent the sets of real and complex numbers, respectively. 
$\mathbb{E}\{ \cdot \}$  is the expectation, and $\Re\{.\}$  returns the real part to a complex number.

%% file: Sec_Model.tex
\section{System Model and Problem Formulation}
\subsection{System Model}
\begin{figure}
	\centering
	\includegraphics[width=1\columnwidth]{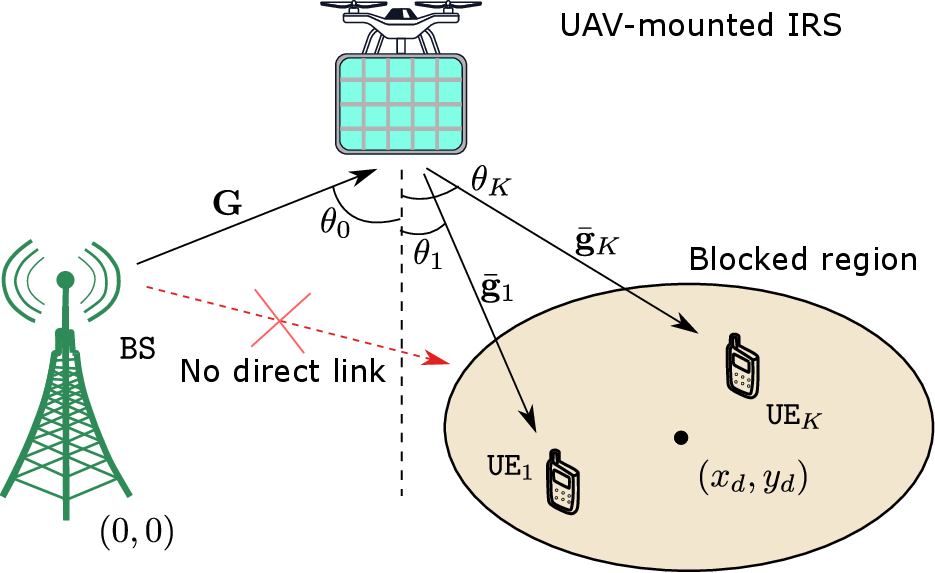}
	\caption{Illustration of the UAV-mounted IRS DL system.}
	\label{fig:system_model}
\end{figure}



We consider a DL system, as shown in Fig. \ref{fig:system_model}, where a BS equipped with $ N $ antennas serves $ K $ single-antenna UEs. Suppose that direct links between the BS and UE are obstructed, and an IRS mounted on the UAV, which comprises $ M $ elements, is deployed to assist the system via the reflecting link.  We denote the set of UE and IRS elements by $ \calK \triangleq \{1,2,\dots,K\} $ and $ \calM \triangleq \{1,2,\dots,M\} $, respectively. Without loss of generality, we consider the 3D Cartesian coordinates and assume that the coordinates of the BS, $ \UEk $, and IRS are $ \mathbf{u}_{\BS} = [x_{\BS},y_\BS,h_\BS]^T, \mathbf{u}_{\UE,k} = [x_{\UE,k},y_{\UE,k},0]^T $ and $ \mathbf{u}_{\UAV} = [x_{\UAV},y_{\UAV},\hUAV]^T $, respectively. Herein, $ \mathbf{u}_{\BS} $ and $ \mathbf{u}_{\UE,k} $ are assumed to be the coordinates assigned during the time interval. Subsequently, the UAV placement is executed by computing the variables $ \mathbf{u}_{\UAV} $ using an optimization algorithm. 

To accelerate the application of a UAV-mounted IRS, we further consider the radiation pattern of the IRS which is modeled as in \cite{IRS_channel_model}, i.e.,
\begin{IEEEeqnarray}{ll} \label{eq: rad. pat.}
	F(\theta,\varphi) = \begin{cases}
		\cos^3(\theta), & \theta \in [0,\pi/2], \varphi \in [0,2\pi] \\
		0, & \theta \in (\pi/2 , \pi] , \varphi \in [0,2\pi],
	\end{cases}
\end{IEEEeqnarray}
where $ \theta $ and $ \varphi $ are the elevation and azimuth angles from the IRS to the BS/UE direction. It is noteworthy that the radiation pattern of the IRS is constant across different azimuth angles. 
For simplicity, we omit the argument $\varphi$ of function $ F(\theta,\varphi) $ in \eqref{eq: rad. pat.}, hereafter, i.e., we use $ F(\theta) $ instead of $ F(\theta,\varphi) $. Therefore, the channels between the BS and IRS, as well as between IRS and $ \UEk $ can be modeled as 
\begin{IEEEeqnarray}{ll}
	\barG   &=\bG \sqrt{F(\theta_0) \beta_0 },  	\non \\
	\barg_k &= \bg_k \sqrt{F(\theta_k) \beta_k }. 		\non
\end{IEEEeqnarray}
Herein, $ \beta_0$ and $ \beta_k$ denote the path loss, whereas $\bG \in \mathbb{C}^{M \times N} $ and $ \bg_k  \in \mathbb{C}^{M \times 1}$ denote the small-scale fading of links $ \BS - \UAV $ and $\UAV- \UEk $, respectively. Note that the small-scale fading is considered to be static during each coherence interval, while the path loss changes much more slowly, which is reasonable since the distances between the users, the BS, and the UAV are much larger than the distance between the BS antennas and between the IRS elements \cite{rappaBook}. 
Accordingly, $ \beta_0 $ and $ \beta_k $ are defined as 
\begin{IEEEeqnarray}{ll}\label{eq:PL_model}
	\beta_{k'} \triangleq c_0 \| \bd_{k'} \|^{-\alpha_{k'}}, \; k'\in\mathcal{K} \cup \{0\}, 
\end{IEEEeqnarray}
where $ c_0 $ is the reference path loss at a distance of $ 1 $m. $ \alpha_0 $ and $ \alpha_{k},\;\forall k\in\mathcal{K} $ are the path loss exponents of links $ \BS - \UAV $ and $\UAV- \UEk $, respectively, whereas $ \bd_0 = \mathbf{u}_{\UAV} - \mathbf{u}_{\BS} = [x_{\UAV} - x_{\BS} , y_{\UAV} - y_{\BS} , \hUAV - \hBS]^T$ and $ \bd_k = \mathbf{u}_{\UAV} - \mathbf{u}_{\UE,k} = [x_{\UAV} - x_{\UE,k} , y_{\UAV} - y_{\UE,k} , \hUAV ]^T$ are the distance vectors from the UAV to the BS and $ \UEk $, respectively.

Finally, the signal received at $ \UEk $ can be written as
\begin{IEEEeqnarray}{ll}\label{eq:y_k}
	y_k = \underset{\forall \ell \in \calK}{\sum} \barg_k^H \bPhi \barG \bw_\ell + n_k,
\end{IEEEeqnarray}
where $ \bw_{\ell} \in \mathbb{C}^{N \times 1}$ is the beamforming vector for $ \UEell $, and $ n_{k} \sim \mathcal{CN}(0,\sigma_{k}^2) $ is the additive white Gaussian noise at $ \UEk $. Meanwhile, $ \bPhi \triangleq \mathtt{diag}( \phi_1, \phi_2, \dots , \phi_M ) $ is the IRS phase shift matrix, in which $ \phi_m \in \mathbb{C} $ denotes the reflecting coefficient of the $ m- $th IRS element, i.e., $ \phi_m \triangleq \beta_{m}^{\IRS} e^{j \theta_{m}^{\IRS}} $ with $ \beta_{m}^{\IRS} \in [0,1] $ and $ \theta_{m}^{\IRS} \in [0,2 \pi]$ being the reflecting amplitude and phase shift of the $ m- $th IRS element, respectively. Here, $ j $ in the exponent term of $ \phi_m $ represents the imaginary unit. Let $ \mathbf{w} = \left[\mathbf{w}_{1}^{T} \mathbf{w}_{2}^{T} \cdots \mathbf{w}_{K}^{T}\right], \btheta \triangleq [\theta_0,\theta_k,\dots,\theta_K] $ and $ \bbeta \triangleq [\beta_0,\beta_1,\dots,\beta_K] $; therefore, the signal-to-interference-plus-noise-ratio (SINR) at $ \UEk $ can be expressed as shown in (4) (top of the next page).
\begin{figure*}
\begin{align} \notag
	\gamma_k(\bw,\bPhi,\btheta,\bbeta) = \frac{ F(\theta_k) F(\theta_0) |\bg_k^H \bPhi \bG \bw_k|^2 \beta_k \beta_0}{ \sum\nolimits_{\forall \ell \in \calK \backslash \{k\}} F(\theta_k) F(\theta_0) |\bg_k^H \bPhi \bG \bw_\ell|^2 \beta_k \beta_0 + \sigma_k^2}.
\label{eq:SINR} \tag{4}
\end{align}
\hrule
\end{figure*}

\subsection{Problem Formulation}
Based on the SINR expression in \eqref{eq:SINR}, the SE in units of nats/s/Hz at $ \UEk $ is written as
\begin{IEEEeqnarray}{ll}\label{eq:SE_UEk}
\setcounter{equation}{5}
	R(\gamma_k(\bw,\bPhi,\btheta,\bbeta)) = \ln (1 + \gamma_k(\bw,\bPhi,\btheta,\bbeta) ),
\end{IEEEeqnarray}
hence, the sum throughput can be expressed as
\begin{IEEEeqnarray}{ll}\label{eq:sumSE}
	\Fsr(\bw,\bPhi,\btheta,\bbeta) = B \sum_{\forall k \in \calK} R(\gamma_k(\bw,\bPhi,\btheta,\bbeta)),
\end{IEEEeqnarray}
where $ B $ is the system bandwidth.

In this study, we aim to maximize the sum throughput by jointly optimizing the beamforming, IRS phase shift, and UAV position based on per-UE QoS requirements. Hence, the optimization problem is formulated as  \footnote{For the sake of formulating a manageable optimization problem, the UAV is assumed to operate without any jitter under the ideal flight condition, and to have sufficient power during its flight-time.}
\begin{IEEEeqnarray}{cl}\label{eq:Prob_maxSE}
	\IEEEyesnumber
	\max_{\substack{\bw,\bPhi,\btheta,\bbeta}} \quad & \Fsr(\bw,\bPhi,\btheta,\bbeta),  \subnum	\label{eq:Prob_maxSE_a}	\\
	\subjectto&  \sum_{\forall k \in \calK}{\|\bw_{k}\|^2 \leq \Pmax }, \subnum  \label{eq:Prob_maxSE_b} \\
	& R(\gamma_k(\bw,\bPhi,\btheta,\bbeta)) \geq \bar{R}_k / B, \forall k \in \calK,		\subnum	\label{eq:Prob_maxSE_c} \\
	& |\phi_m| \leq 1, \forall m \in \calM. \subnum \label{eq:Prob_maxSE_d} 
\end{IEEEeqnarray}
Here, constraint \eqref{eq:Prob_maxSE_b} guarantees that the total transmitted power does not exceed the maximum power at the BS, \eqref{eq:Prob_maxSE_c} represents the QoS requirement for each UE, and \eqref{eq:Prob_maxSE_d} describes the reflecting coefficient constraint of each IRS element. It is clear that the optimization problem \eqref{eq:Prob_maxSE} is nonconvex owing to the non-concave objective function \eqref{eq:Prob_maxSE_a} and non-convex constraint \eqref{eq:Prob_maxSE_c}. Therefore, it is generally difficult to solve this problem directly.

%% file: Sec_Proposed_SR.tex
\section{Proposed IA-based algorithm}
In this section, we propose an iterative algorithm that employs BCD and IA methods. In particular, the beamforming and IRS phase shift are alternately optimized with the UAV positions by BCD, in which the non-convex objective and constraint are addressed using the IA method. Hence, the original problem can be decomposed into two subproblems, wherein the UAV position and either beamforming or IRS phase shift are jointly optimized in each sub-problem. Before deriving a more tractable form for problem \eqref{eq:Prob_maxSE}, we first introduce the IA-based approximation functions that will be applied at each iteration $i$}:
\begin{itemize}
	\item A lower bound of the convex function $ |x|^2/y, x \in \mathbb{C}, y \geq 0 $ can be obtained as \cite{HoangTuy:Globaloptimization}
	\begin{IEEEeqnarray}{ll}\label{eq:bound_x2/y}
		\frac{|x|^2}{y} \geq 2\frac{(x^{(i)})^H x}{y^{(i)}} - \frac{|x^{(i)}|^2}{(y^{(i)})^2}y.
	\end{IEEEeqnarray}
	\item Considering the multiplication function $ \fmul(x,y) \triangleq xy, x \geq 0, y\geq 0$, the upper bound of $ \fmul(x,y) $ is presented as
	\begin{IEEEeqnarray}{ll} \label{eq:multiply function}
		\fmul(x,y) \leq \frac{ y^{(i)} }{ 2 x^{(i)}} x^2 + \frac{ x^{(i)} }{ 2 y^{(i)} } y^2 := \fmul^{(i)}(x,y).
	\end{IEEEeqnarray}
	\item Define the quadratic function $ \fqu(\mathbf{x}) \triangleq \| \mathbf{x} \|^2 , \mathbf{x} \in \mathbb{C}^L, L \in \mathbb{R}, L \geq 1 $, then its lower bound is given as in \cite{DinhJSAC18}, i.e.,
	\begin{IEEEeqnarray}{ll} \label{eq:lower bound quadractic}
		\fqu(\mathbf{x}) \geq 2 \mathfrak{R}\{ ( \mathbf{x}^{(i)} )^H \mathbf{x} \} - \| \mathbf{x}^{(i)} \|^2 := \fqu^{(i)}(\mathbf{x}).
	\end{IEEEeqnarray}
	\item The lower bound of the convex function $ \fpow(x;a) \triangleq  x^a, \; x \geq 0, a \geq 1 $ is expressed as
	\begin{IEEEeqnarray}{ll} \label{eq:lower bound x^a}
		\fpow(x; a) & \; \geq a ( x^{(i)} )^{a-1} x  - (a-1)(x^{(i)})^a \nonumber  \\
		& \; := \fpow^{(i)}(x; a). \qquad
	\end{IEEEeqnarray}
\end{itemize}

To ease the difficulty of non-convexity in the original problem \eqref{eq:Prob_maxSE}, the following lemma is introduced to derive a more tractable form.

\begin{mylem}
	We first set $ \blambda \triangleq \{ \lambda_k \}_{\forall k \in \calK} $ and $ \bLambda \triangleq \mathtt{diag}(\lambda_1,\lambda_2,\dots,\lambda_K) $ as the smooth variables for the SINRs and then apply the IA method to the distance-based attenuation terms of the SINR function in \eqref{eq:SINR}. Without loss of optimality, an approximated optimization problem for solving \eqref{eq:Prob_maxSE} at iteration $ i $ of an iterative algorithm is formulated as
	\begin{IEEEeqnarray}{cl}\label{eq:Prob_maxSE2}
		\IEEEyesnumber
		\max_{\substack{\bw,\bPhi,\calU,\blambda}} \quad & \Fsrbar(\bLambda) \triangleq B \ln | \mathbf{I}_K + \bLambda | \subnum	\label{eq:Prob_maxSE2_a}	\\
		\subjectto&  \bar{\gamma}_k(\bw,\bPhi,\bzeta_{\Upper},\bzeta_{\Lower}, \bmu) \geq \lambda_k, \forall k \in \calK,		\subnum	\label{eq:Prob_maxSE2_b} \\
		& \lambda_k + 1 \geq \exp(\bar{R}_k / B), \forall k \in \calK, \subnum	\label{eq:Prob_maxSE2_c} \\
		& \|\mathbf{d}_{k'} \|^{\bar{\alpha}_{k'}} \leq \rho_{\Upper,k'}, \; \forall k'\in\calK\cup\{0\},  \subnum \label{eq:Prob_maxSE2_d1} \\
		& \fmul^{(i)}( \rho_{\Upper,0}, \rho_{\Upper,k} ) \leq \zeta_{\Upper,k}, \; \forall k\in\calK,\subnum \label{eq:Prob_maxSE2_d2}  \\
		& \tilde{\rho}_{\Lower,k'} \leq \fqu^{(i)}(\mathbf{d}_{k'}), \forall k' \in \calK\cup \{0\}, \subnum \label{eq:Prob_maxSE2_d3}  \\
		& \rho_{\Lower,k'} \leq \fpow^{(i)}(\tilde{\rho}_{\Lower,k'}; \bar{\alpha}_{k'}/2), \forall k' \in \calK\cup \{0\}, \subnum \qquad \label{eq:Prob_maxSE2_d4}  \\
		& \bar{\rho}_{\Lower,k}^2 \leq \rho_{\Lower,0} \rho_{\Lower,k}, \; \forall k\in\calK, \subnum \label{eq:Prob_maxSE2_d5}  \\
		& \zeta_{\Lower,k} \leq \fqu^{(i)}(\bar{\rho}_{\Lower,k}^2 ), \; \forall k\in\calK, \subnum \label{eq:Prob_maxSE2_d6}  \\
		& \mu_k \geq \frac{\sigma_k^2}{\fqu(\upsilon)},  \; k\in \calK,\subnum \label{eq:Prob_maxSE2_d7} \\ 
		& \upsilon^2 \leq  \fpow(\hUAV;3) \fpow(\hUAV-\hBS;3), \subnum \label{eq:Prob_maxSE2_d8} \\
		& \eqref{eq:Prob_maxSE_b},\eqref{eq:Prob_maxSE_d},  \subnum \label{eq:Prob_maxSE2_d}
	\end{IEEEeqnarray}
	where $ \bar{\alpha}_{k'} \triangleq 3 + \alpha_{k'},\;\forall k'\in\calK\cup\{0\} $, and the SINR at $ \UEk $ is given as
	\begin{IEEEeqnarray}{ll}
		\bar{\gamma}_k(\bw,\bPhi,\bzeta_{\Upper},\bzeta_{\Lower},\bmu) \triangleq \frac{ c_0^2 |\bg_k^H \bPhi \bG \bw_k|^2 \zeta_{\Upper,k}^{-1}  }
		{ \underset{\forall \ell \neq k \in \calK }{\sum} c_0^2 |\bg_k^H \bPhi \bG \bw_\ell|^2 \zeta_{\Lower,k}^{-1} + \mu_k }. \non
	\end{IEEEeqnarray}
	We set a variable $ \calU \triangleq \{\buUAV,\brho_{\Upper},\bzeta_{\Upper},\brho_{\Lower},\tilde{\brho}_{\Lower},\bar{\brho}_{\Lower},\bzeta_{\Lower}, \break \bmu, \upsilon\} $, with the new variables defined as
	\begin{align}
		\brho_\Upper & \triangleq \{\rho_{\Upper,k'}\}, \; {\forall k' \in \calK\cup\{0\}}, \nonumber \\
		\bzeta_{\Upper} & \triangleq \{\zeta_{\Upper,k}\}, \; {\forall k \in \calK}, \nonumber \\
		\brho_\Lower & \triangleq \{\rho_{\Lower,k'}\}, \; {\forall k' \in \calK\cup\{0\}}, \nonumber \\
		\tilde{\brho}_\Lower & \triangleq \{\tilde{\rho}_{\Lower,k'}\}, \; {\forall k' \in \calK\cup\{0\}}, \nonumber \\
		\bar{\brho}_\Lower & \triangleq \{\bar{\rho}_{\Lower,k}\}, \; {\forall k \in \calK}, \nonumber \\
		\bzeta_{\Lower} & \triangleq \{\zeta_{\Lower,k}\}, \; {\forall k \in \calK}, \nonumber \\
		\bmu & \triangleq \{\mu_k\}, \; {\forall k\in\calK}. \nonumber
	\end{align}

\end{mylem}

\begin{IEEEproof}
	Please see Appendix. 
\end{IEEEproof}

Clearly, the objective function is expressed in a logdet form that is a concave function. In addition, the QoS constraint \eqref{eq:Prob_maxSE_c} is rewritten in linear form \eqref{eq:Prob_maxSE2_c}, with respect to the new variable $ \blambda $. {However, constraint \mbox{\eqref{eq:Prob_maxSE2_b}} remains nonconvex owing to the combination of optimization variables, including the phase shift matrix, beamforming vectors, UAV position and auxiliary variables (e.g., $ \bzeta_{\Upper} $ and $ \bzeta_{\Lower} $ are used to convexify the constraints of the radiation pattern and UAV location).} Next, we propose a BCD-based method to address the product of the phase shift matrix and beamforming vectors, while they are jointly optimized with the other variables.
The main notations are summarized in Table~\ref{tab:notation}.

\begin{table}[t]
	\caption{List of main notations}
	\label{tab:notation}
	\centering
		\scalebox{0.85}{
		\begin{tabular}{c|l}
			\hline
			Notation & Description \\
			\hline\hline
			$ \barG $	& Channel gain matrix IRS -- BS \\
			$ \barg_k $	& Channel gain vector IRS -- $ \UEk $ \\
			$ \bw_k $ 	& Beamforming vector for $ \UEk $ \\
			$ \bPhi $	& Phase shift matrix of IRS \\
			$ \btheta $   &  Elevation angle of the boresight direction IRS -- BSs/UEs \\
			$ \bd_0 $	& Distance vector IRS--BS \\
			$ \bd_k $	& Distance vector IRS--$ \UEk $ \\
			$ \brho_\Upper $ &	An upper bound of power function of distance between IRS and BS/UE \\
			$ \bzeta_{\Upper} $ & Upper bound of product of $ \brho_\Upper $ \\
			$ \brho_\Lower $ &	Lower bound of power function of distance between IRS and BS/UE \\
			$ \bzeta_{\Lower} $ & Lower bound of product of $ \brho_\Lower $ \\
			$ \bmu $	& Upper bound of noise function \\
			$ \blambda $	& Lower bound of SINR function \\
			$ \btau $	& Upper bound of interference power in SINR function \\
			$ \bomega $	& Lower bound of signal power in SINR function \\
			\hline		   				
		\end{tabular}
		}
\end{table}

\subsection{Updating of $ \bw $ and $ \buUAV $}
In this subsection, we formulate the subproblem wherein the beamforming and UAV position are jointly optimized with a fixed IRS phase shift. Based on  \eqref{eq:Prob_maxSE2}, the subproblem around a feasible point $ (\bw^{(i)},\calU^{(i)}|\bar{\bPhi}) $ at iteration $ i $ can be derived  as
\begin{IEEEeqnarray}{cl}\label{eq:Prob_maxSE_updateW}
	\IEEEyesnumber
	\max_{\substack{\bw,\calU,\blambda}} \quad & \Fsrbar(\bLambda) \subnum	\label{eq:Prob_maxSE_updateW_a}	\\
	\subjectto&  \bar{\gamma}_k(\bw,\bzeta_{\Upper},\bzeta_{\Lower},\bmu|\bar{\bPhi}) \geq \lambda_k, \forall k \in \calK,	\subnum	\label{eq:Prob_maxSE_updateW_b} \\
	& \eqref{eq:Prob_maxSE_b}, \eqref{eq:Prob_maxSE2_c}-\eqref{eq:Prob_maxSE2_d8}, \subnum \label{eq:Prob_maxSE_updateW_c}
\end{IEEEeqnarray}
where $ \bar{\gamma}_k(\bw,\bzeta_{\Upper},\bzeta_{\Lower},\bmu|\bar{\bPhi}) $ is given by $ \bar{\gamma}_k(\bw,\bPhi,\bzeta_{\Upper},\bzeta_{\Lower},\bmu) $ with $ \bPhi $ being fixed to $ \bar{\bPhi} $. It is clear that \eqref{eq:Prob_maxSE_updateW_b} is still nonconvex.

To convexify \eqref{eq:Prob_maxSE_updateW_b}, we first introduce a new variable $ \btau=\{\tau_{k,\ell}\}_{\forall k,\ell \in \calK} $ as the upper bound of the interference power in the denominator of $ \bar{\gamma}_k(\bw,\bzeta_{\Upper},\bzeta_{\Lower}|\bar{\bPhi}) $, which satisfies the following rotated cone constraint:
\begin{IEEEeqnarray}{lll} \label{eq:bound interference updateW}
	\frac{|\barg_k^H \bar{\bPhi} \barG \bw_\ell|^2}  {\zeta_{\Lower,k}} \leq \tau_{k,\ell}, \forall k,\ell \in \calK, k\neq \ell.
\end{IEEEeqnarray}
Next, the numerator of $ \bar{\gamma}_k(\bw,\bzeta_{\Upper},\bzeta_{\Lower}|,\bar{\bPhi}) $ is lower-bounded as
\begin{IEEEeqnarray}{ll} \label{eq:bound signal updateW}
	\frac{|\bar{g}_k^H \bar{\bPhi} \bG \bw_k|^2}{ \zeta_{\Upper,k} }	\geq \omega_{k}^2, \; \forall k \in \calK,
\end{IEEEeqnarray}
where $ \bomega \triangleq \{\omega_{k}\}_{\forall k \in \calK} $ is a new variable. By applying \eqref{eq:bound_x2/y} to the left-hand side of \eqref{eq:bound signal updateW}, we obtain the following convex constraint: 
\begin{IEEEeqnarray}{ll} \label{eq:bound signal updateW - convex}
	\fsk (\bw,\bar{\bPhi},\bzeta_{\Upper}; \bw^{(i)}, \bar{\bPhi}) \geq \omega_{k}^2, \; \forall k \in \calK,
\end{IEEEeqnarray}
where
\begin{IEEEeqnarray}{ll}
	&\fsk (\bw,\bPhi,\bzeta_{\Upper}; \bw^{(i)}, \bPhi^{(i)} ) \triangleq \nonumber \\
	& \qquad \qquad \qquad \qquad 2 \frac{\mathfrak{R} ((\bar{g}_k^H \bPhi^{(i)} \bG \bw_k^{(i)})*(\bar{g}_k^H \bPhi \bG \bw_k))}{\zeta_{\Upper,k}^{(i)}} \nonumber \\
	& \qquad \qquad \qquad \qquad - \frac{ |\bar{g}_k^H \bPhi^{(i)} \bG \bw_k^{(i)}|^2 }{ (\zeta_{\Upper,k}^{(i)})^2 } \zeta_{\Upper,k}.
\end{IEEEeqnarray}

Based on \mbox{\eqref{eq:bound_x2/y}}, it is shown that $ \bar{\gamma}_k(\bw,\bzeta_{\Upper},\bzeta_{\Lower}|,\bar{\bPhi}) $ has a lower bound as
\begin{IEEEeqnarray}{ll}\label{eq:convex_SINR_updateW}
	\bar{\gamma}_k(\bw,\bzeta_{\Upper},\bzeta_{\Lower},\bmu|,\bar{\bPhi}) &  \geq \frac{c_0^2 \omega_k^2}{\Psi_k(\btau,\bmu)} \nonumber \\
	& \geq \calA_{1,k}(\bomega) - \calA_{2,k}(\btau,\bmu), \quad
\end{IEEEeqnarray}
where
\begin{IEEEeqnarray}{ll}
	 \Psi_k(\btau,\bmu) & \triangleq \underset{\forall \ell \in \calK \backslash \{k\} }{\sum} c_0^2 \tau_{k,\ell} + \mu_k, \nonumber \\
	 \Psi_k^{(i)}  & \triangleq \Psi_k(\btau^{(i)},\bmu^{(i)}), \non \\
	\calA_{1,k}(\bomega) & \triangleq 2 \frac{c_0^2 \omega_k^{(i)}*\omega_k }{ \Psi_k^{(i)} },	\non  \\
	\calA_{2,k}(\btau,\bmu) & \triangleq \frac{ c_0^2 (\omega_{k}^{(i)})^2 }{ (\Psi_k^{(i)})^2 } \underset{\forall \ell \neq k \in \calK }{\sum}  c_0^2 \tau_{k,\ell} + \mu_k \non.
\end{IEEEeqnarray}

The subproblem for joint beamforming and UAV position optimization at iteration $ i+1 $ can be expressed as
\begin{IEEEeqnarray}{cl}\label{eq:Prob_maxSE_updateW2}
	\IEEEyesnumber
	\max_{\substack{\bw,\calU,\bomega,\btau,\blambda}} \quad & \Fsrbar(\bLambda) \subnum	\label{eq:Prob_maxSE_updateW2_a}	\\
	\subjectto
	& \calA_{1,k}(\bomega) - \calA_{2,k}(\btau,\bmu) \geq \lambda_k, \;k\in\calK, \subnum \label{eq:Prob_maxSE_updateW2_b} \\
	&  \eqref{eq:Prob_maxSE_b}, \eqref{eq:Prob_maxSE2_c}-\eqref{eq:Prob_maxSE2_d8}, \eqref{eq:bound interference updateW}, \eqref{eq:bound signal updateW - convex}.  \subnum \label{eq:Prob_maxSE_updateW2_c} 
\end{IEEEeqnarray}

\begin{algorithm}[t]
	\begin{algorithmic}[1]
		\protect\caption{Proposed Algorithm to Solve Problem \eqref{eq:Prob_maxSE}}
		\label{alg_1}
		\STATE{\textbf{Initialization:}} Set $i:=0$, $ \epsilon=10^{-3} $ and generate the initial point $(\bw^{(0)},\bPhi^{(0)},\calU^{(0)})$ by solving \eqref{eq:Prob get initial point}.\\
		\WHILE {$ \frac{| c_{\bw}-c_{\bPhi} |}{B} \geq \epsilon $}
		\REPEAT
		\STATE Solve \eqref{eq:Prob_maxSE_updateW2} to find $(\bw^\star,\calU^\star, \bomega^\star, \btau^\star)$.
		\STATE Update $(\bw^{(i+1)},\calU^{(i+1)}, \bomega^{(i+1)}, \btau^{(i+1)}):=(\bw^\star,\calU^\star, \bomega^\star, \btau^\star)$.
		\STATE Set $i:=i+1.$
		\UNTIL Convergence 
		\STATE Set $ c_{\bw} = \Fsrbar(\bLambda) $
		\REPEAT
		\STATE Solve \eqref{eq:Prob_maxSE_updatePhi2} to find $(\bPhi^\star,\calU^\star, \bomega^\star, \btau^\star)$.
		\STATE Update $(\bPhi^{(i+1)},\calU^{(i+1)}, \bomega^{(i+1)}, \btau^{(i+1)}):=(\bPhi^\star,\calU^\star, \bomega^\star, \btau^\star)$.
		\STATE Set $i:=i+1.$
		\UNTIL Convergence 
		\STATE Set $ c_{\bPhi} = \Fsrbar(\bLambda) $.
		\ENDWHILE
		\STATE \textbf{Output:} The optimal solution $ (\bw^\star, \bPhi^\star , \buUAV^\star) $.
\end{algorithmic} \end{algorithm}

\subsection{Updating of $ \bPhi $ and $ \buUAV $}
In this subsection, we jointly optimize the IRS phase shift and UAV position while fixing the beamforming. Therefore, the sub-problem at feasible point $ (\bPhi^{(i)},\calU^{(i)}|\bar{\bw}) $ with fixed $ \bar{\bw} $ can be expressed as follows:
\begin{IEEEeqnarray}{cl}\label{eq:Prob_maxSE_updatePhi}
	\IEEEyesnumber
	\max_{\substack{\bPhi,\calU,\blambda}} \quad & \Fsrbar(\bLambda) \subnum	\label{eq:Prob_maxSE_updatePhi_a}	\\
	\subjectto&  \bar{\gamma}_k(\bPhi,\bzeta_{\Upper},\bzeta_{\Lower}|\bar{\bw}) \geq \lambda_k, \forall k \in \calK,		\subnum	\label{eq:Prob_maxSE_updatePhi_b} \\
	& \eqref{eq:Prob_maxSE_b}, \eqref{eq:Prob_maxSE2_c}-\eqref{eq:Prob_maxSE2_d6}. \subnum \label{eq:Prob_maxSE_updatePhi_c}
\end{IEEEeqnarray}
Constraint \eqref{eq:Prob_maxSE_updatePhi_b} can be convexified using the steps presented in \eqref{eq:Prob_maxSE_updateW_b}. First, to optimize $ \bPhi $ instead of $ \bw $, constraints \eqref{eq:bound interference updateW} and \eqref{eq:bound signal updateW - convex} are updated as
\begin{IEEEeqnarray}{rl} \label{eq:bound interference updatePhi}
	\IEEEyesnumber
	\frac{|\barg_k^H \bPhi \barG \bar{\bw}_\ell|^2}{\zeta_{\Lower,k}} & \leq \tau_{k,\ell}, \forall k,\ell \in \calK, k\neq \ell, \subnum \\
	\fsk (\bar{\bw},\bPhi,\bzeta_{\Upper}; \bar{\bw}, \bar{\bPhi}^{(i)}) & \geq \omega_{k}^2, \; \forall k \in \calK. \subnum
\end{IEEEeqnarray}
Similarly to \mbox{\eqref{eq:convex_SINR_updateW}}, the lower bound of $ \bar{\gamma}_k(\bPhi,\bzeta_{\Upper},\bzeta_{\Lower}|\bar{\bw}) $ is expressed as
\begin{IEEEeqnarray}{ll}\label{eq:convex_SINR_updatePhi}
	\bar{\gamma}_k(\bPhi,\bzeta_{\Upper},\bzeta_{\Lower},\bmu|\bar{\bw}) & \geq \frac{\omega_k^2}{\Psi_k(\btau,\bmu)} \non \\
	& \geq \calA_{1,k}(\bomega) - \calA_{2,k}(\btau,\bmu). \quad
\end{IEEEeqnarray}
Hence, the solution for \eqref{eq:Prob_maxSE_updatePhi} can be obtained by solving the following subproblem at iteration $ i+1 $:
\begin{IEEEeqnarray}{cl}\label{eq:Prob_maxSE_updatePhi2}
	\IEEEyesnumber
	\max_{\substack{\bPhi,\calU,\bomega,\btau,\blambda}} \quad & \Fsrbar(\bLambda) \subnum	\label{eq:Prob_maxSE_updatePhi2_a}	\\
	\subjectto&  \calA_{1,k}(\bomega) - \calA_{2,k}(\btau,\bmu) \geq \lambda_k, \forall k \in \calK, \quad	\subnum \\
	& \eqref{eq:Prob_maxSE_d}, \eqref{eq:Prob_maxSE2_c}-\eqref{eq:Prob_maxSE2_d8},\eqref{eq:bound interference updatePhi}.  \subnum \label{eq:Prob_maxSE_updatePhi2_b}
\end{IEEEeqnarray}

\subsection{Generation of an initial point} 
To execute either \eqref{eq:Prob_maxSE_updateW2} or \eqref{eq:Prob_maxSE_updatePhi2}, an initial point $ (\bw^{(0)},\bPhi^{(0)},\calU^{(0)}) $ is necessary. In this study, subproblem \eqref{eq:Prob_maxSE_updateW2} is solved with a first iteration, where $ \bPhi^{(0)} $ is generated by a random phase shift with a reflecting amplitude of $ 1 $, and the initial point for beamforming and UAV position are obtained by satisfying the QoS requirement at the fixed IRS phase shift. Hence, we solve the following subproblem to obtain the initial point for \eqref{eq:Prob_maxSE_updateW2}:
\begin{IEEEeqnarray}{cl}\label{eq:Prob get initial point}
	\IEEEyesnumber
	\max_{\substack{\bw,\calU,\bomega,\btau,\blambda,\bdelta}} \quad & \Delta \triangleq \sum_{\forall k \in \calK}{\delta_k} \subnum	\label{eq:Prob get initial point a}	\\
	\subjectto& \lambda_k + 1 - \exp(\bar{R}_k / B) \geq \delta_k, \forall k \in \calK,	\qquad	\subnum	\label{eq:Prob get initial point b} \\
	& \delta_k \leq 0, \forall k \in \calK,		\subnum	\label{eq:Prob get initial point c} \\
	& \eqref{eq:Prob_maxSE_b}, \eqref{eq:Prob_maxSE2_d1}-\eqref{eq:Prob_maxSE2_d8}, \eqref{eq:bound interference updateW}, \eqref{eq:bound signal updateW - convex}, \eqref{eq:Prob_maxSE_updateW2_b} \subnum \label{eq:Prob get initial point d}
\end{IEEEeqnarray}
where $ \bdelta \triangleq \{\delta_k\}_{\forall k \in \calK} $ is a new variable. An initial point for \eqref{eq:Prob_maxSE_updateW2} is obtained when objective $ \Delta $ is approximately zero. Finally, the proposed algorithm is summarized in Algorithm 1.

\subsection{Complexity analysis}
To compute the complexity of Algorithm 1, we first determine the complexity of the two subproblems based on the number of constraints and variables. The number of cone/linear constraints of sub-problems \mbox{\eqref{eq:Prob_maxSE_updateW2}} and \mbox{\eqref{eq:Prob_maxSE_updatePhi2}} are the same, i.e., $ c = (K^2 + 9K + 5 )$. The number of variables in \mbox{\eqref{eq:Prob_maxSE_updateW2}} is $ v_1 = (K^2 + 9K + NK + 7) $, whereas that in \mbox{\eqref{eq:Prob_maxSE_updatePhi2}} is $ v_2 = (K^2 + 9K + MN + 7) $. According to \cite{sedumi}, the per-iteration complexities for solving \eqref{eq:Prob_maxSE_updateW2} and \eqref{eq:Prob_maxSE_updatePhi2} in terms of the big-O can be expressed as $ \mathcal{O} (v_1^{2.5} (c + v_1)) $ and $ \mathcal{O} (v_2^{2.5} (c + v_2)) $, respectively. Therefore, the per-iteration of Algorithm 1 is equivalent to $ \mathcal{O} (v_1^{2.5} (c + v_1)) + \mathcal{O} (v_2^{2.5} (c + v_2)) $, which is comparable to the complexity $ \mathcal{O} (v_1^{2.5} (c + v_1)) $ of the method without phase shift optimization.

%% file: Sec_Results.tex
\section{Numerical Results}
In this section, we evaluate the effectiveness of the proposed approach for the considered UAV-mounted IRS system, where the system parameters for the simulation are set as follows:
\begin{itemize}
\item The BS is equipped with $N=16$ antennas located at $ (0,0,30)$, i.e., $ \hBS = 30 $ m with a power budget $ \Pmax = 38 $ dBm.
\item $K = 6$ single-antenna UEs are assumed to be uniformly distributed in a circular area with radius $ R_{\mathtt{d}} = 30$ m and center position $ (x_{\mathtt{d}},y_{\mathtt{d}}) = (0,50)$.
\item The UAV altitude $\hUAV$ is within a range $ [30, 120] $ m.
\item The number of passive elements in the IRS is $ M=50 $.
\item The QoS requirements of all UEs, without loss of generality, are set to $ \bar{R}_k = \bar{R} = 5$ Mbits, $ \forall k \in \calK $.
\item The path-loss exponents for BS-UAV and UAV-UE links are set to $ \alphaBSUAV = 2 $ and $ \alphaUAVUE = 2.2 $, and the reference path-loss is set to $ c_0 = -30 $ dB for all links \cite{Ge_Access_20}.
\item The $ K $ factors for the Rician channels BS-UAV and UAV-UE links are set to $ K_{\BS-\UAV} = 10 $ dB and $ K_{\UAV-\UE} = 5 $ dB, respectively.
\end{itemize}

\begin{figure}
	\centering
	\includegraphics[width=1\linewidth]{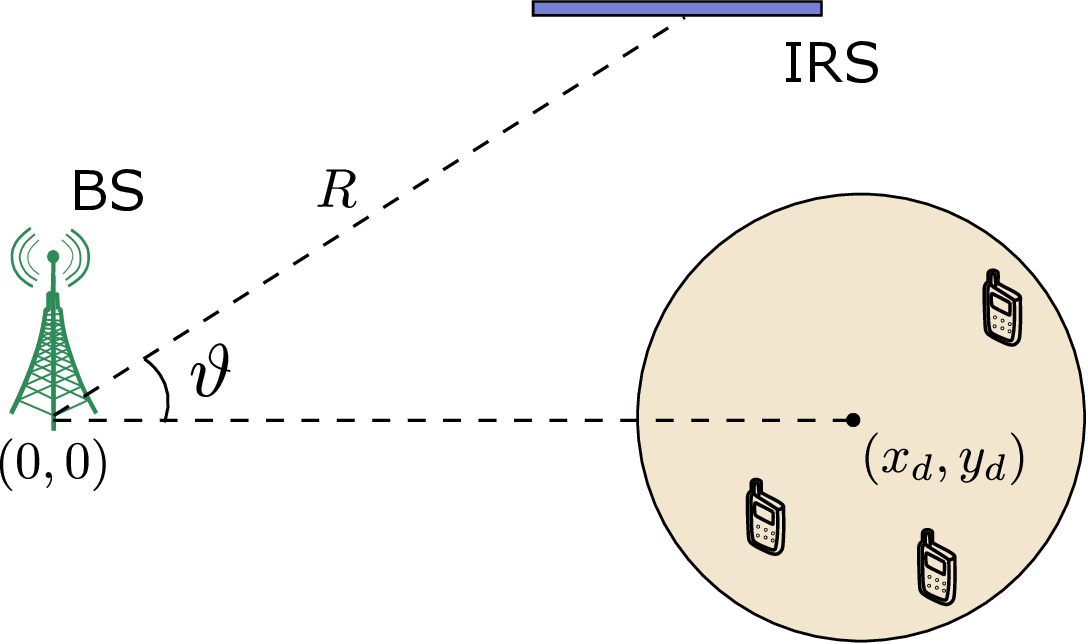}
	\caption{Simulation setup for TIRS scenario.}
	\label{fig:ground_IRS}
\end{figure}

\begin{figure}
	\centering
	\includegraphics[width=1\linewidth]{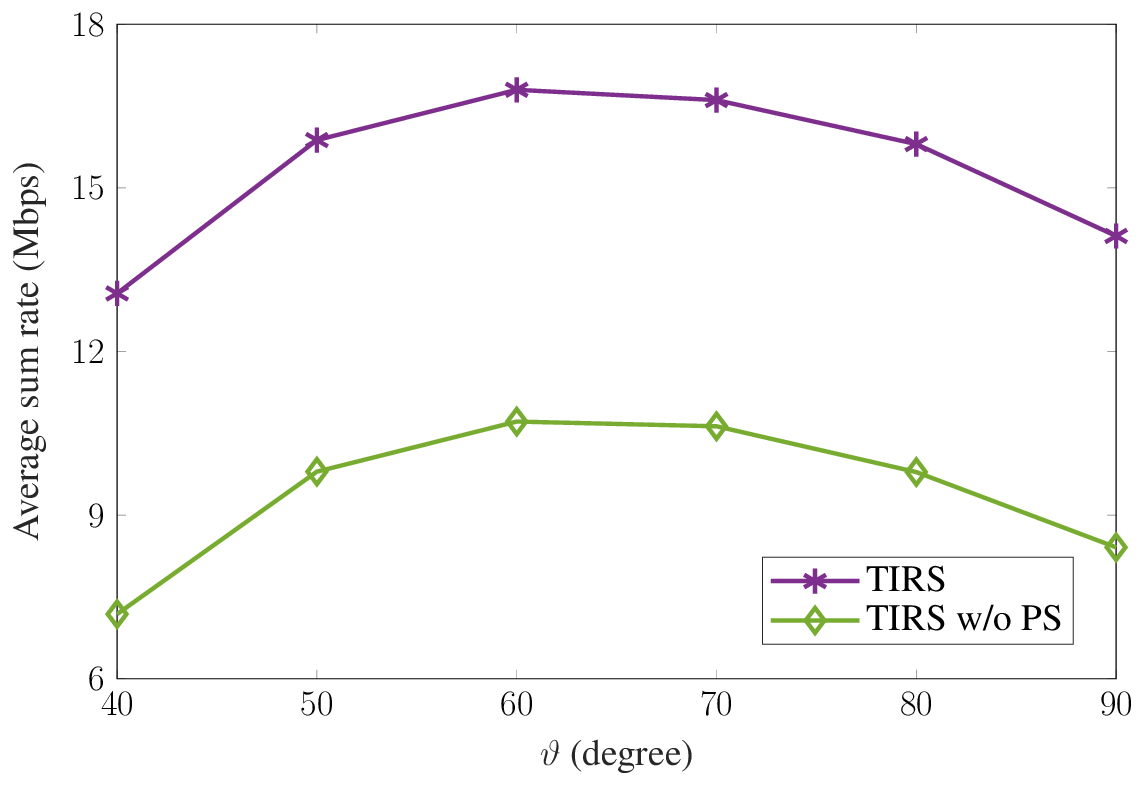}
	\caption{The average sum rate of TIRS scenario with difference values of $ \vartheta $.}
	\label{fig:SR_groundIRS_angle}
\end{figure}

For comparison purpose, we consider the same simulation setup for the terrestrial IRS (TIRS) counterpart as that for the proposed UAV-mounted IRS model, as illustrated in Fig. \mbox{\ref{fig:ground_IRS}}. In particular, we assume that the IRS is placed on a building at the same altitude as the BS, and the path-loss exponents for BS-IRS and IRS-UE links are set to $ 2 $ and $ 2.4 $, respectively \mbox{\cite{WCNC_IRS_GreenNetwork,UAV_IRS_Symbiotic_System}}. To achieve the best performance in the case of TIRS, we examine the relative position of the IRS based on angle $ \vartheta $, which is displayed in Fig. \mbox{\ref{fig:ground_IRS}}. Accordingly, Fig. \mbox{\ref{fig:SR_groundIRS_angle}} shows the average sum rate with different values of $ \vartheta $ in the TIRS scenario. As shown, both schemes with and without phase shift optimization yield the best sum rate with $ \vartheta = 60^{\circ} $. Therefore, we select $ \vartheta = 60 ^{\circ} $ for the TIRS scenario.

Next, to evaluate the effectiveness of the proposed UAV-mounted IRS (UmI) approach with optimized altitude, referred to as ``UmI w/ $\hUAV$ optimized'' scheme, we compare its performance with that of the following four schemes:
\begin{itemize}
	\item UmI: In this scheme, UAV is deployed at the fixed altitude $ \hUAV = 70 $ m, i.e., only the horizontal placement of the UAV, the beamforming at the BS, and phase shift at the IRS are jointly optimized.
	\item UmI without (w/o) phase shift (PS): Only horizontal placement of UmI and BS beamforming are jointly optimized in this scheme, while the IRS phase shift matrix $ \bPhi $ is set to an identity matrix.
	\item TIRS: As described in Fig. 2, the terrestrial IRS is placed on a building at the same altitude as the BS, considering the phase shift optimization. Herein, the beamforming at the BS and the IRS phase shift are alternately optimized.
	\item TIRS w/o PS: This scheme is similar to the TIRS scheme, but only the beamforming at the BS is optimized. The phase shift matrix is set to an identity matrix.
\end{itemize}

\begin{figure}
	\centering
	\includegraphics[width=1\linewidth]{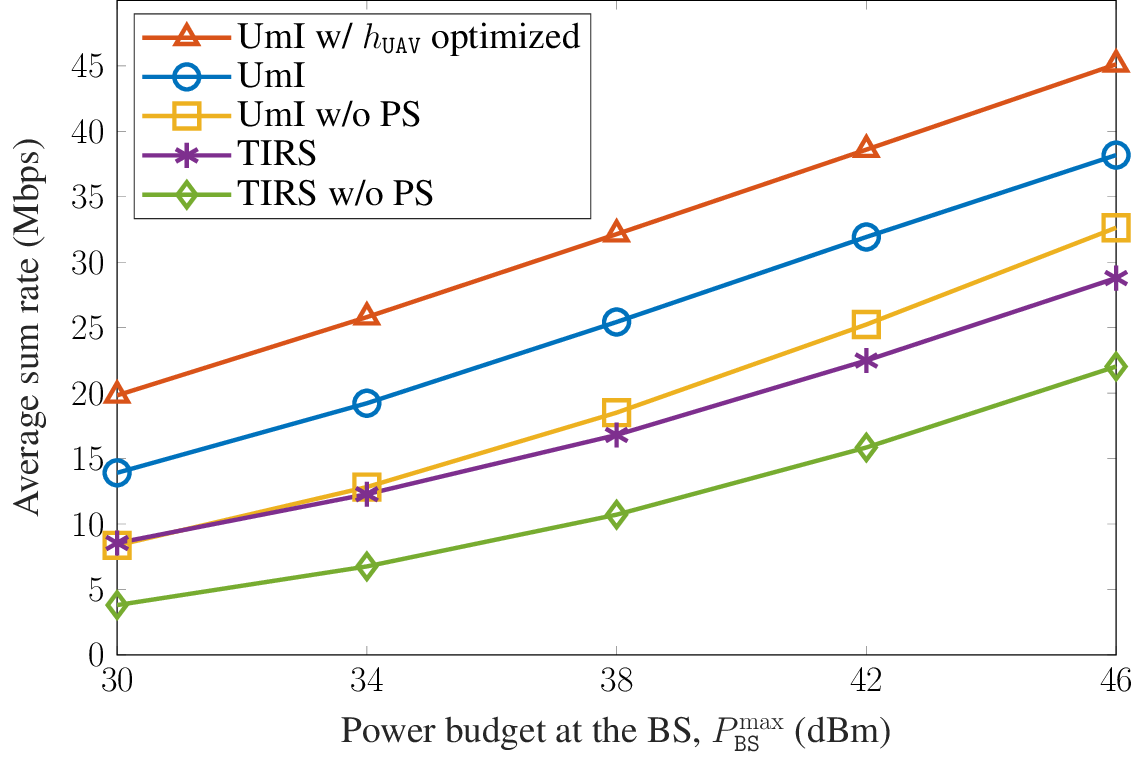}
	\caption{The average sum rate (Mbps) as a function of the power budget at the BS.}
	\label{fig:SR_Pmax}
\end{figure}

Fig. \mbox{\ref{fig:SR_Pmax}} shows the average sum rate performance (in Mbps) with respect to the maximum power levels at the BS, $ \Pmax $ (in dBm). First, it is observe that the sum rates of all considered schemes increase rapidly with the BS maximum power, whereas the performance gain in the average sum rate of the UmI over TIRS remains constant, i.e., approximately $ 9 $ Mbps. As expected, the phase shift  optimization brings significant improvement in sum rate for both the TIRS and UmI scenarios, especially for UmI cases. For example, at $ \Pmax = 38 $ dBm, the sum rate of UmI and TIRS cases upon applying phase shift optimization grows by about $ 14 $ Mbps and 6 Mbps, respectively (from 18.5 Mbps to 32.1 Mbps for UmI and from 10.7 Mpbs to 16.8 Mbps for TIRS). Moreover, by mounting the IRS on the UAV platform, the sum rate gap between the proposed UmI system with optimized altitude and the TIRS counterpart enlarges as $\Pmax$ increases, from about $ 10 $ Mpbs at $\Pmax = 30 $ dBm to about $ 23 $ Mbps at $\Pmax = 46 $ dBm. In the end, for the UmI model, the altitude optimization can bring substantial sum rate enhancement, compared to the fixed altitude case, with approximately $ 6 $ Mbps at all considered $ \Pmax $ values, making the scheme UmI w/ $\hUAV$ optimized superior to all the rests.

\begin{figure}
	\centering
	\includegraphics[width=1\linewidth]{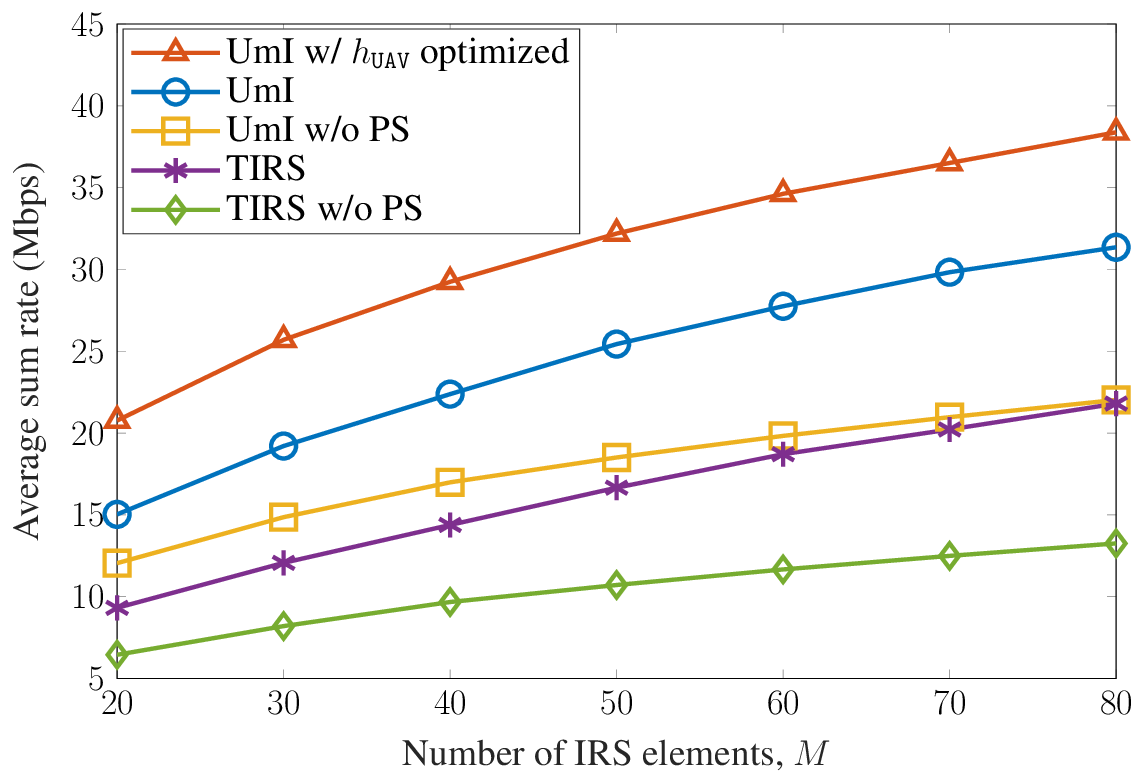}
	\caption{The average sum rate versus the number of IRS elements.}
	\label{fig:SR_NumIRSelements}
\end{figure}

In Fig. \mbox{\ref{fig:SR_NumIRSelements}}, the average sum rate of all schemes is determined as a function of the number of IRS elements, $M$. Here, similar sum rate behaviors can be observed as in Fig. \mbox{\ref{fig:SR_Pmax}}, which confirms the benefits of the employed IRS-integrated UAV system and the effectiveness of the phase shift optimization. By providing the IRS with additional DoF by attaching it to UAV, the sum rate gap between the proposed UmI model, even with fixed UAV altitude, and the TIRS case increases twice, when the number of IRS elements increases from 20 to 80, i.e., from 5 Mbps at $ M = 20 $ to $ 10 $ Mbps at $ M = 80 $. This performance gap between the UmI and TIRS models appears more remarkable when the UAV altitude is optimized. For instance, the sum rate of UmI with optimized $\hUAV$ reaches approximately $ 38.3 $ Mbps with $ M = 80$ IRS elements, whereas that of the UmI with fixed $\hUAV$ and TIRS is about $ 31.3 $ Mbps and $ 21.8 $ Mbps, respectively.

\begin{figure}
	\centering
	\includegraphics[width=1\linewidth]{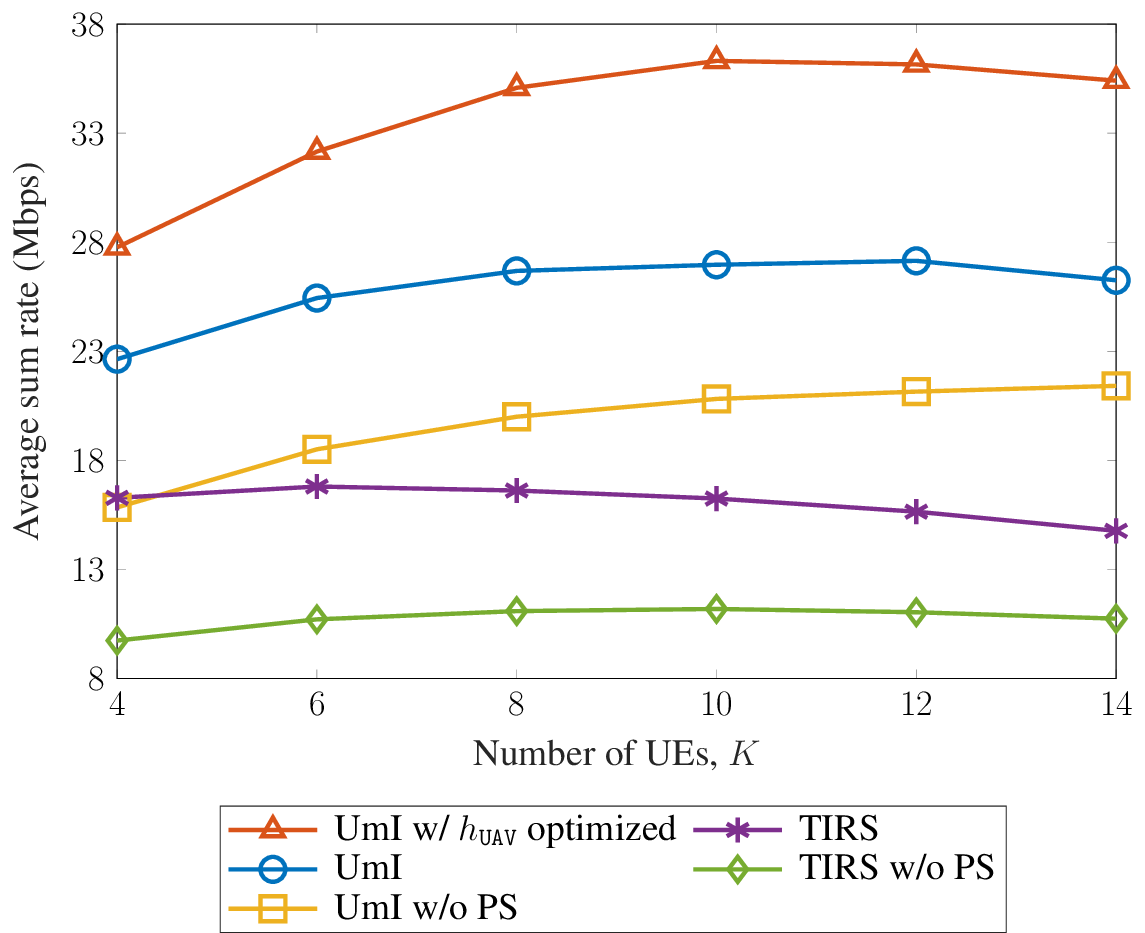}
	\caption{The average sum rate with the different number of UEs, $ K $.}
	\label{fig:SR_NumUEs}
\end{figure}

Next, we investigate the effect of the number of UEs, $K$, on the system sum rate, as shown in Fig. \mbox{\ref{fig:SR_NumUEs}}. It is interesting to see that the gain afforded by the phase shift optimization of some schemes tends to decrease as the number of UEs increases. This occurs because the DoF provided by the IRS relative to the number of UEs is reduced when $ K $ grows. In addition, the sum rate curves of all five schemes exhibit a similar trend: The sum rate first increases and then decreases, as the number of UEs increases. This reflects the fact that, for a relatively small number of UEs, the DoF and maximum power at the BS are sufficient for improving the sum rate performance. However, when a significant number of UEs are involved, the DoF and $ \Pmax $ become deficient for the sum rate enhancement. On the other hand, the SR performance gap between UmI and UmI w/ $ \hUAV $ optimized becomes larger as the number of UEs increases. This result confirms the effectiveness of 3D UAV placement optimization (with $ \hUAV $ optimized) compared with horizontal UAV placement optimization (with fixed $ \hUAV $) for a large number of UEs.

\begin{figure}
	\centering
	\includegraphics[width=1\linewidth]{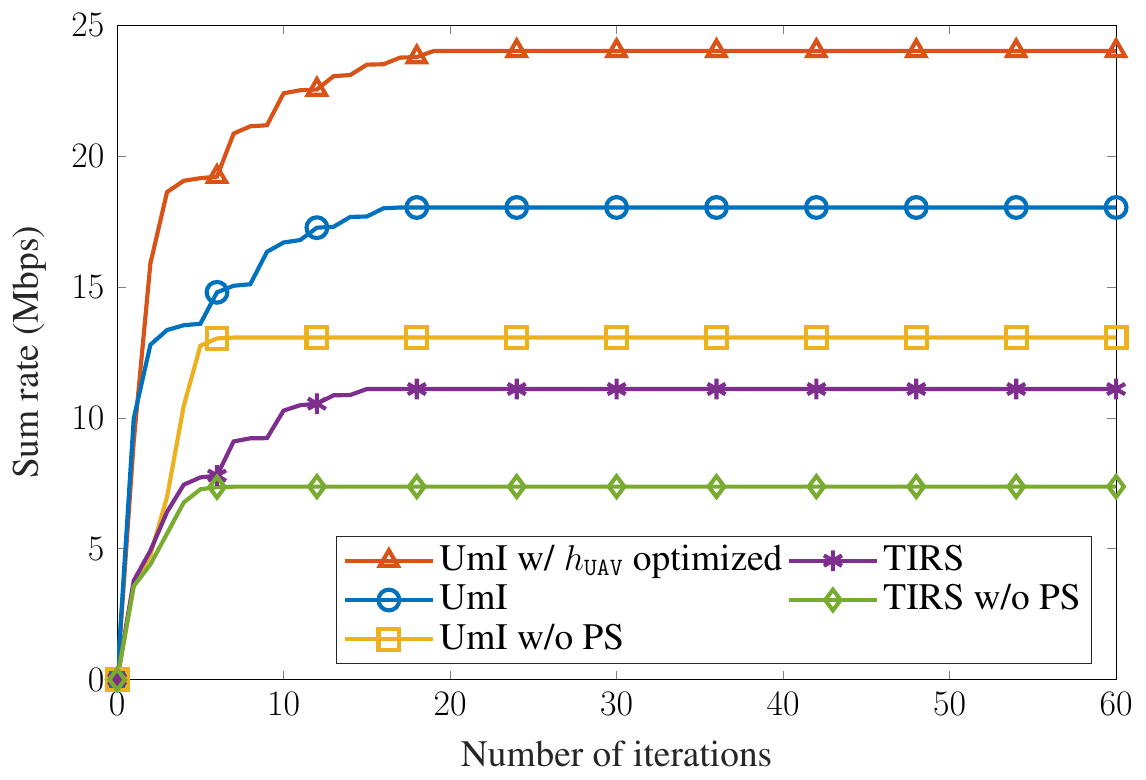}
	\caption{The convergence behavior of the sum rate performance.}
	\label{fig:SR_Convergence}
\end{figure}

Another important performance metric that needs to be investigated is the convergence rate of different schemes. Here we set  $ \Pmax = 34$ dBm, $M=50,N = 16 $, and $ K=6 $. Fig. \mbox{\ref{fig:SR_Convergence}} displays the sum rates obtained in all schemes as a function of the number of iterations. Explicitly, the sum rate of both the UmI and TIRS cases w/o PS quickly converges, i.e., approximately $ 10 $ iterations to achieve $ 99\% $ performance. This happens because only the beamforming is computed at the BS. Meanwhile, the optimization of the IRS phase shift requires more operating time to reach the saturated level (at about $ 16 $ iterations); however, it provides a considerable improvement of the sum rate for both the UmI and TIRS cases (about $45\% $ performance gain). Furthermore, it revealed that the jumping steps in the convergence behavior of both the UmI and TIRS schemes are caused by the alternating optimization procedure of the BS beamforming and the IRS phase shift. Accordingly, this reduces the convergence speed in the two above-mentioned cases upon phase shift optimization procedure. Nevertheless, such convergence rate performance is reasonably compensated by the considerable sum rate gains afforded by adjusting the phase shift.

\begin{figure}
	\centering
	\includegraphics[width=1\linewidth]{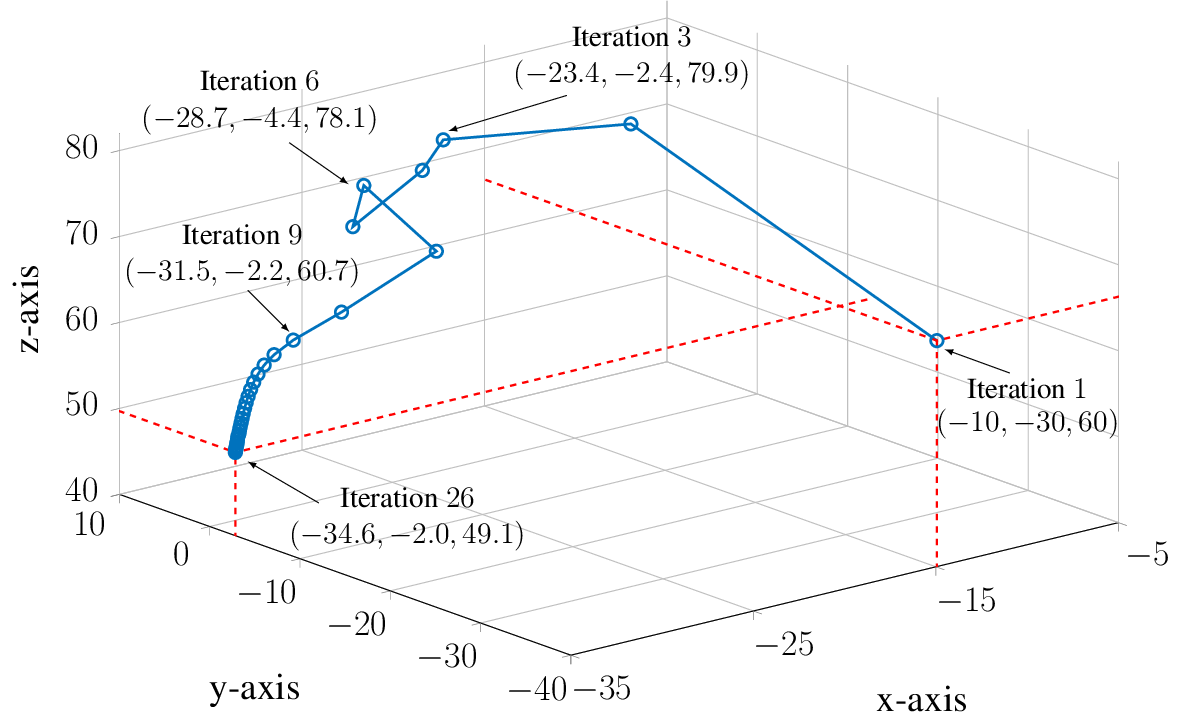}
	\caption{UAV placement convergence behavior.}
	\label{fig:uav_placement_convergence}
\end{figure}

As a matter of fact, UmI w/ $\hUAV $ optimized always achieves the best performance among all the investigated schemes and keeps a large constant gain over the counterpart scheme with the fixed altitude, i.e., UmI. A rising question is how fast the optimization problem of UmI w/ $\hUAV $ optimized converges compared to the other schemes. To answer the question, Fig. \mbox{\ref{fig:uav_placement_convergence}} shows the convergence behavior of the UAV placement for the UmI w/ $\hUAV$ optimized. It is seen that it requires only about ten iterations to reach nearly convergence point, which validates the effectiveness of the proposed algorithm. 
Another interesting point is that the altitude $\hUAV$ converges to about $ 49 $ m in the range $ [30, 120] $ m. The convergence point is not close to the lower limit, because it depends on both the radiation pattern of IRS and pathloss of the channel. 

\begin{figure}
	\centering
	\includegraphics[width=1\linewidth]{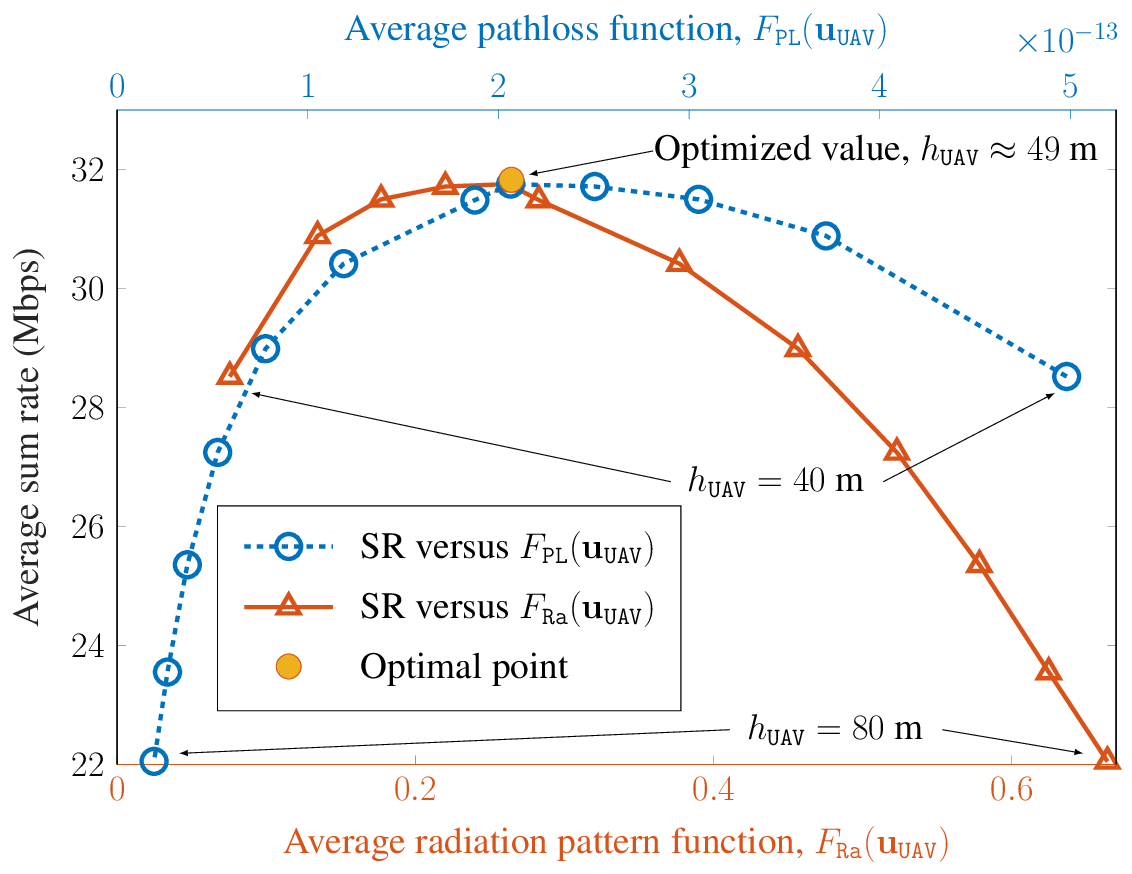}
	\caption{Trade-off between radiation pattern and pathloss functions.}
	\label{fig:tradeoff_hUAV}
\end{figure}

To gain a deep insight into this behavior, we further investigate a trade-off between the radiation pattern and pathloss, which impacts on the average sum rate, as shown in  Fig. \mbox{\ref{fig:tradeoff_hUAV}}. 
Note that both the average pathloss $\FPL(\buUAV)$ and the average radiation pattern $\FRa(\buUAV)$ can be defined as functions of $\hUAV$, i.e., $ \FPL(\buUAV) \triangleq \big( \sum_{\forall k \in \calK} \| \bd_0\|^{-\alpha_{0}} \| \bd_k \|^{-\alpha_{k}} \big) /  K $ and $ \FRa(\buUAV) \triangleq \big(\sum_{\forall k \in \calK} (\hUAV - \hBS)^3 \hUAV^3 {(\| \bd_0\| \| \bd_k \|)}^{-3} \big) /  K$. Interestingly, prior to the optimal altitude ( $ \hUAV^{\mathtt{opt}} = 49$ m), the sum rate increases in proportion to the growth of $ \FPL(\buUAV) $ and $ \FRa(\buUAV) $, but in the opposite direction of $ \hUAV $ variation. More specifically, the sum rate as a function of $ \FRa(\buUAV) $ decreases from $\hUAV^{\mathtt{opt}} = 49$ m to $\hUAV =$80 m, while the sum rate as a function of $\FPL(\buUAV) $ decreases from $\hUAV^{\mathtt{opt}} = 49$ m to $\hUAV =$40 m.

%% file: Sec_Conclusion.tex
\section{Conclusion}

In this work, we investigated a UAV-mounted IRS system to assist the DL communication between the multi-antenna BS and multiple UEs, wherein the BS beamforming, IRS phase shift and UAV placement were jointly optimized, targeting the sum rate maximization. To address the nonconvexity of the original problem, we employed both the BCD and IA methods to segregate the problem into two sub-problems, and then we developed an iterative algorithm to alternately optimize the BS beamforming and IRS phase shift with UAV placement. Numerical results demonstrated that our proposed approach for UAV-aided aerial IRS DL system brings significant sum rate enhancement compared to the conventional system with the terrestrial IRS counterpart, even with the fixed altitude of the UAV. Moreover, it was shown that the sum rate gap becomes more remarkable when the UAV altitude is allowed to vary in the proposed optimization problem.

%% file: Sec_Appendix.tex
\appendix
\section*{Appendix}

To relieve the difficulty of solving problem \eqref{eq:Prob_maxSE}, we transform \eqref{eq:Prob_maxSE} into a more tractable form. First, we introduce a new variable $ \blambda \triangleq \{ \lambda_k \}_{\forall k \in \calK}$ to smooth the SINR function:
\begin{IEEEeqnarray}{ll} \label{eq:smooth SINR}
	\gamma_k(\bw,\bPhi,\btheta,\bbeta) \geq \lambda_k, \forall k \in \calK.
\end{IEEEeqnarray}
Subsequently, the lower bound of the objective function \eqref{eq:Prob_maxSE_a} is rewritten as
\begin{IEEEeqnarray}{ll} \label{eq:transformed_obj}
	\Fsr(\bw,\btheta,\bvarphi,\bPhi,\bbeta) &\geq B \sum_{\forall k \in \calK} R(\lambda_k) = B \ln \big( \prod_{\forall k \in \calK}(1 + \lambda_k) \big)	\non \\
	& = B \ln | \mathbf{I}_K + \bLambda |  \triangleq \Fsrbar(\bLambda),
\end{IEEEeqnarray}
where $ \mathbf{I}_K $ is the $K\times K$ unit matrix and $ \bLambda \triangleq \mathtt{diag}(\lambda_1,\lambda_2,\dots,\lambda_K) $. Clearly, \eqref{eq:transformed_obj} is performed as a concave logdet function. However, the additional constraint \eqref{eq:smooth SINR} remains highly non-convex; hence, the remainder of the proof is to derive a more solvable form for \eqref{eq:smooth SINR}.

Let us extract the radiation pattern of the IRS in \eqref{eq: rad. pat.} as a function with respect to the distance from the UAV to the BS and to the UEs, i.e.,
\begin{IEEEeqnarray}{ll}\label{eq:transformed_radiation_pattern}
	F(\theta_0) =  \frac{(\hUAV - h_{\BS})^3}{\| \bd_0 \|^3}, \subnum \\
	F(\theta_k) = \frac{\hUAV^3}{\| \bd_k \|^3}. \subnum 
\end{IEEEeqnarray}
By substituting \mbox{\eqref{eq:PL_model}} and \mbox{\eqref{eq:transformed_radiation_pattern}} into \mbox{\eqref{eq:SINR}}, $ \gamma_k(\bw,\bPhi,\btheta,\bbeta),\;k\in\calK $ in \mbox{\eqref{eq:smooth SINR}} can be rewritten as
\begin{IEEEeqnarray}{ll}\label{eq:SINR_equi}
	\begin{aligned}
		& \gamma_k(\bw,\bPhi,\buUAV) = \\
		&  \frac{ c_0^2 |\bg_k^H \bPhi \bG \bw_k|^2 \|\bd_0 \|^{-\bar{\alpha}_{0}} \| \bd_k \|^{-\bar{\alpha}_{k}}  }
		{ \underset{\forall \ell \neq k \in \calK }{\sum} c_0^2 |\bg_k^H \bPhi \bG \bw_\ell|^2 \|\bd_0 \|^{-\bar{\alpha}_{0}} \|\bd_k\|^{-\bar{\alpha}_{k}} + \fnk(\hUAV)}. \qquad
	\end{aligned}
\end{IEEEeqnarray}
where $ \fnk(\hUAV) = \frac{\sigma_k^2}{\hUAV^3 (\hUAV - h_{\BS})^3} $.

We further address the norm functions of $ \gamma_k(\bw,\bPhi,\buUAV) $ based on the linear upper and lower bounds of $ \|\mathbf{d}_0 \|^{\bar{\alpha}_{0}} \|\mathbf{d}_k \|^{\bar{\alpha}_{k}} $. The upper bound of $ \|\mathbf{d}_0 \|^{\bar{\alpha}_{0}} \|\mathbf{d}_k \|^{\bar{\alpha}_{k}} $, denoted by $ \zeta_{\Upper,k} $, around point $ \buUAV^{(i)} $ at iteration $ i $ of an iterative algorithm is obtained as the following constraint:
\begin{IEEEeqnarray}{ll} \label{eq:upper bound d^a leq zeta}
	\|\mathbf{d}_0 \|^{\bar{\alpha}_{0}} \|\mathbf{d}_k \|^{\bar{\alpha}_{k}} \leq \zeta_{\Upper,k},\; \forall k \in \calK. \quad
\end{IEEEeqnarray}
To address the non-convexity of \eqref{eq:upper bound d^a leq zeta}, we first introduce a new variable $ \rho_{\Upper,k'} $ and impose the following power-cone constraint:
\begin{IEEEeqnarray}{ll}\label{eq:upper bound d^a leq rho - PCone_PL}
	\|\mathbf{d}_k \|^{\bar{\alpha}_{k'}} \leq \rho_{\Upper,k'},\;\forall k' \in \calK\cup \{0\},
\end{IEEEeqnarray}
hence, constraint \eqref{eq:upper bound d^a leq zeta} becomes 
\begin{IEEEeqnarray}{ll} \label{eq:upper bound d^a leq zeta - aux. var.}
	\rho_{\Upper,0} \rho_{\Upper,k} \leq \zeta_{\Upper,k},\; \forall k \in \calK. \quad
\end{IEEEeqnarray}
By applying \eqref{eq:multiply function} to the left-hand side of \eqref{eq:upper bound d^a leq zeta - aux. var.}, constraint \eqref{eq:upper bound d^a leq zeta} is convexified as
\begin{IEEEeqnarray}{ll} \label{eq:upper bound d^a leq zeta - convex}
	\fmul^{(i)}( \rho_{\Upper,0}, \rho_{\Upper,k} ) \leq \zeta_{\Upper,k},\; \forall k \in \calK. \quad
\end{IEEEeqnarray}

However, the function $ \|\mathbf{d}_0 \|^{\bar{\alpha}_{0}} \|\mathbf{d}_k \|^{\bar{\alpha}_{k}} $ is lower bounded by $ \zeta_{\Lower,k} $, which is expressed as
\begin{align} \label{eq:lower bound zeta leq product of d^a}
	\zeta_{\Lower,k} \leq \|\mathbf{d}_0 \|^{\bar{\alpha}_{0}} \|\mathbf{d}_k \|^{\bar{\alpha}_{k}},\; k \in \calK.
\end{align}
By introducing the new variables $ \{\rho_{\Lower,k'}\}_{\forall k' \in \calK \cup \{0\}} $, constraint \eqref{eq:lower bound zeta leq product of d^a} can be equivalently expressed as
\begin{subnumcases}{}
	\rho_{\Lower,k'} \leq \|\mathbf{d}_{k'} \|^{\bar{\alpha}_{k'}}, \; \forall k' \in \calK \cup \{0\}, \label{eq: lowerbound a} \\
	\zeta_{\Lower,k} \leq \rho_{\Lower,0} \rho_{\Lower,k},\; k \in \calK. \label{eq: lowerbound b}
\end{subnumcases}
Using \eqref{eq:lower bound quadractic} and \eqref{eq:lower bound x^a}, constraint \eqref{eq: lowerbound a} can be approximated as the following convex constraints:
\begin{IEEEeqnarray}{ll}\label{eq:lower bound rho leq d^a}
	\tilde{\rho}_{\Lower,k'} \leq \fqu^{(i)}(\mathbf{d}_{k'}), \forall k' \in \calK \cup \{0\}, \subnum \\
	\rho_{\Lower,k'} \leq \fpow^{(i)}(\tilde{\rho}_{\Lower,k'};\bar{\alpha}_{k'}/2),\; \forall k' \in \calK \cup \{0\}, \subnum
\end{IEEEeqnarray}
whereas constraint \eqref{eq: lowerbound b} is convexfied as
\begin{IEEEeqnarray}{ll}\label{eq:lower bound zeta leq d^a}
	\bar{\rho}_{\Lower,k}^2 \leq \rho_{\Lower,0} \rho_{\Lower,k},\; k \in \calK, \subnum \\
	\zeta_{\Lower,k} \leq \fqu^{(i)}(\bar{\rho}_{\Lower,k}^2 ),\; k \in \calK, \subnum 
\end{IEEEeqnarray}
where $ \{\tilde{\rho}_{\Lower,1}\}, \; \forall k' \in \calK \cup \{0\} $, and $ \{\bar{\rho}_{\Lower,k}\},\; k \in \calK $ are newly introduced as smooth variables.

%
%

Consequently, the lower bound of the SINR in \mbox{\eqref{eq:SINR_equi}} is obtained as
\begin{align}\label{eq:SINR_with_bounded_distance}
	\gamma_k(\bw,\bPhi,\buUAV) & \geq \frac{ c_0^2 |\bg_k^H \bPhi \bG \bw_k|^2 \zeta_{\Upper,k}^{-1}  }
	{ \underset{\forall \ell \neq k \in \calK }{\sum} c_0^2 |\bg_k^H \bPhi \bG \bw_\ell|^2 \zeta_{\Lower,k}^{-1} + \fnk(\hUAV)}.
\end{align}
Now, we let $ \bmu\triangleq\{\mu_k\}_{k\in\calK} $ be an upper bound of $ f_N(\hUAV) \triangleq \{\fnk(\hUAV)\}_{k\in\calK} $, which is expressed as
	\begin{IEEEeqnarray}{ll}\label{eq:upper bound fN hUAV}
		\mu_k \geq \fnk(\hUAV), \; k\in \calK.
	\end{IEEEeqnarray}
Constraint \mbox{\eqref{eq:upper bound fN hUAV}} can be convexified as
\begin{align} \label{eq:upper bound fN hUAV1}
	\mu_k \geq \frac{\sigma_k^2}{\fqu(\upsilon)},  \; k\in \calK,
\end{align}
with the following convex constraints imposed:
\begin{align} \label{eq:upper bound fN hUAV2}
	\upsilon^2 \leq  \fpow(\hUAV;3) \fpow(\rUAVBS;3), \; k\in \calK,
\end{align}
where $ \fpow(\rUAVBS;3) \triangleq \fpow(\hUAV - \hBS;3) $

%
	
Then, instead of $ \tilde{\gamma}_k(\bw,\bPhi,\bzeta_{\Upper},\bzeta_{\Lower}) $, we get a more tractable form of SINR at  $ \UEk $ as
	\begin{IEEEeqnarray}{ll}
		\bar{\gamma}_k(\bw,\bPhi,\bzeta_{\Upper},\bzeta_{\Lower},\bmu) \triangleq \frac{ c_0^2 |\bg_k^H \bPhi \bG \bw_k|^2 \zeta_{\Upper,k}^{-1}  }
		{ \underset{\forall \ell \neq k \in \calK }{\sum} c_0^2 |\bg_k^H \bPhi \bG \bw_\ell|^2 \zeta_{\Lower,k}^{-1} + \mu_k }. \non
	\end{IEEEeqnarray}

Using \mbox{\eqref{eq:transformed_obj}, \eqref{eq:upper bound d^a leq rho - PCone_PL}, \eqref{eq:upper bound d^a leq zeta - convex}, \eqref{eq:lower bound rho leq d^a}, \eqref{eq:upper bound fN hUAV1}, and \eqref{eq:upper bound fN hUAV2}}, we obtain problem \mbox{\eqref{eq:Prob_maxSE2}}, which provides a maximization minorant for the original problem \mbox{\eqref{eq:Prob_maxSE}}.


%% file: IRS_UAV_flexible_hUAV_v5_clean.bbl
\begin{thebibliography}{10}
\providecommand{\url}[1]{#1}
\csname url@samestyle\endcsname
\providecommand{\newblock}{\relax}
\providecommand{\bibinfo}[2]{#2}
\providecommand{\BIBentrySTDinterwordspacing}{\spaceskip=0pt\relax}
\providecommand{\BIBentryALTinterwordstretchfactor}{4}
\providecommand{\BIBentryALTinterwordspacing}{\spaceskip=\fontdimen2\font plus
\BIBentryALTinterwordstretchfactor\fontdimen3\font minus
  \fontdimen4\font\relax}
\providecommand{\BIBforeignlanguage}[2]{{%
\expandafter\ifx\csname l@#1\endcsname\relax
\typeout{** WARNING: IEEEtran.bst: No hyphenation pattern has been}%
\typeout{** loaded for the language `#1'. Using the pattern for}%
\typeout{** the default language instead.}%
\else
\language=\csname l@#1\endcsname
\fi
#2}}
\providecommand{\BIBdecl}{\relax}
\BIBdecl

\bibitem{Holographic_Surface_6G}
C.~Huang, S.~Hu, G.~C. Alexandropoulos, A.~Zappone, C.~Yuen, R.~Zhang, M.~D.
  Renzo, and M.~Debbah, ``Holographic {MIMO} surfaces for {6G} wireless
  networks: Opportunities, challenges, and trends,'' \emph{IEEE Wireless
  Commun.}, vol.~27, no.~5, pp. 118--125, Oct. 2020.

\bibitem{bjorn19}
E.~Bj{\"o}rnson, {\"O}.~{\"O}zdogan, and E.~G. Larsson, ``Intelligent
  reflecting surface versus decode-and-forward: How large surfaces are needed
  to beat relaying?'' \emph{IEEE Wireless Commun. Lett.}, vol.~9, no.~2, pp.
  244--248, Feb. 2020.

\bibitem{IRS_aided_wireless_networkJ_reconfigurable_environment}
Q.~Wu and R.~Zhang, ``Towards smart and reconfigurable environment: Intelligent
  reflecting surface aided wireless network,'' \emph{IEEE Commun. Mag.},
  vol.~58, no.~1, pp. 106--112, Jan. 2020.

\bibitem{RISvsRelaying}
M.~Di~Renzo, K.~Ntontin, J.~Song, F.~H. Danufane, X.~Qian, F.~Lazarakis,
  J.~De~Rosny, D.-T. Phan-Huy, O.~Simeone, R.~Zhang, M.~Debbah, G.~Lerosey,
  M.~Fink, S.~Tretyakov, and S.~Shamai, ``Reconfigurable intelligent surfaces
  vs. relaying: Differences, similarities, and performance comparison,''
  \emph{IEEE Open J. Commun. Society}, vol.~1, pp. 798--807, June 2020.

\bibitem{hum2013}
S.~V. Hum and J.~Perruisseau-Carrier, ``Reconfigurable reflectarrays and array
  lenses for dynamic antenna beam control: A review,'' \emph{IEEE Trans.
  Antennas Propag.}, vol.~62, no.~1, pp. 183--198, Jan. 2014.

\bibitem{Zeng19}
Y.~Zeng, Q.~Wu, and R.~Zhang, ``Accessing from the sky: A tutorial on {UAV}
  communications for {5G} and beyond,'' \emph{Proc. IEEE}, vol. 107, no.~12,
  pp. 2327--2375, Dec. 2019.

\bibitem{Huo19}
Y.~Huo, X.~Dong, T.~Lu, W.~Xu, and M.~Yuen, ``Distributed and multilayer {UAV}
  networks for next-generation wireless communication and power transfer: A
  feasibility study,'' \emph{IEEE Internet Things J.}, vol.~6, no.~4, pp.
  7103--7115, Aug. 2019.

\bibitem{Mozaffari19}
M.~Mozaffari, W.~Saad, M.~Bennis, Y.-H. Nam, and M.~Debbah, ``{A tutorial on
  UAVs for wireless networks: Applications, challenges, and open problems},''
  \emph{IEEE Commun. Surv. Tut.}, vol.~21, no.~3, pp. 2334--2360, 3rd Quarter
  2019.

\bibitem{Li_WCL_20}
S.~Li, B.~Duo, X.~Yuan, Y.-C. Liang, and M.~Di~Renzo, ``Reconfigurable
  intelligent surface assisted {UAV} communication: Joint trajectory design and
  passive beamforming,'' \emph{IEEE Wireless Commun. Lett.}, vol.~9, no.~5, pp.
  716--720, May 2020.

\bibitem{Wei_TWC_21}
Z.~Wei, Y.~Cai, Z.~Sun, D.~W.~K. Ng, J.~Yuan, M.~Zhou, and L.~Sun, ``Sum-rate
  maximization for {IRS}-assisted {UAV} {OFDMA} communication systems,''
  \emph{IEEE Trans. Wireless Commun.}, vol.~20, no.~4, pp. 2530--2550, Apr.
  2021.

\bibitem{MU_JSAC_21}
X.~Mu, Y.~Liu, L.~Guo, J.~Lin, and H.~Vincent~Poor, ``Intelligent reflecting
  surface enhanced multi-{UAV} {NOMA} networks,'' \emph{IEEE J. Select. Areas
  Commun.}, vol.~39, no.~10, pp. 3051--3066, Oct. 2021.

\bibitem{Ge_Access_20}
L.~Ge, P.~Dong, H.~Zhang, J.-B. Wang, and X.~You, ``Joint beamforming and
  trajectory optimization for intelligent reflecting surfaces-assisted {UAV}
  communications,'' \emph{IEEE Access}, vol.~8, pp. 78\,702--78\,712, Apr.
  2020.

\bibitem{UAV_IRS_Symbiotic_System}
M.~Hua, L.~Yang, Q.~Wu, C.~Pan, C.~Li, and A.~Lee~Swindlehurst,
  ``{UAV}-assisted intelligent reflecting surface symbiotic radio system,''
  \emph{IEEE Trans. Wireless Commun.}, vol.~20, no.~9, pp. 5769--5785, Sept.
  2021.

\bibitem{PangTCOM21}
X.~Pang, N.~Zhao, J.~Tang, C.~Wu, D.~Niyato, and K.-K. Wong, ``{IRS}-assisted
  secure {UAV} transmission via joint trajectory and beamforming design,''
  \emph{IEEE Trans. Commun.}, vol.~70, no.~2, pp. 1140--1152, Feb. 2022.

\bibitem{Sun_TVT_21}
G.~Sun, X.~Tao, N.~Li, and J.~Xu, ``Intelligent reflecting surface and {UAV}
  assisted secrecy communication in millimeter-wave networks,'' \emph{IEEE
  Trans. Veh. Technol.}, vol.~70, no.~11, pp. 11\,949--11\,961, Nov. 2021.

\bibitem{Pan_WCL_21}
Y.~Pan, K.~Wang, C.~Pan, H.~Zhu, and J.~Wang, ``{UAV}-assisted and intelligent
  reflecting surfaces-supported {Terahertz} communications,'' \emph{IEEE
  Wireless Commun. Lett.}, vol.~10, no.~6, pp. 1256--1260, June 2021.

\bibitem{Shafique_TCOM_21}
T.~Shafique, H.~Tabassum, and E.~Hossain, ``Optimization of wireless relaying
  with flexible {UAV}-borne reflecting surfaces,'' \emph{IEEE Trans. Commun.},
  vol.~69, no.~1, pp. 309--325, Jan. 2021.

\bibitem{Mahmoud_TGN_21}
A.~Mahmoud, S.~Muhaidat, P.~Sofotasios, I.~Abualhaol, O.~A. Dobre, and
  H.~Yanikomeroglu, ``Intelligent reflecting surfaces assisted {UAV}
  communications for {IoT} networks: Performance analysis,'' \emph{IEEE Trans.
  Green Commun. Net.}, vol.~5, no.~3, pp. 1029--1040, Sept. 2021.

\bibitem{PangWCL21}
X.~Pang, M.~Sheng, N.~Zhao, J.~Tang, D.~Niyato, and K.-K. Wong, ``When {UAV}
  meets {IRS}: Expanding air-ground networks via passive reflection,''
  \emph{IEEE Wireless Commun.}, vol.~28, no.~5, pp. 164--170, Oct. 2021.

\bibitem{Zhou21}
T.~Zhou, K.~Xu, X.~Xia, W.~Xie, and J.~Xu, ``{Achievable rate optimization for
  aerial intelligent reflecting surface-aided cell-free massive {MIMO}
  system},'' \emph{IEEE Access}, vol.~9, pp. 3828--3837, Dec. 2020.

\bibitem{Lu_TWC_21}
H.~Lu, Y.~Zeng, S.~Jin, and R.~Zhang, ``Aerial intelligent reflecting surface:
  Joint placement and passive beamforming design with {3D} beam flattening,''
  \emph{IEEE Trans. Wireless Commun.}, vol.~20, no.~7, pp. 4128--4143, July
  2021.

\bibitem{liaskos2018}
C.~Liaskos, S.~Nie, A.~Tsioliaridou, A.~Pitsillides, S.~Ioannidis, and
  I.~Akyildiz, ``A new wireless communication paradigm through
  software-controlled metasurfaces,'' \emph{IEEE Commun. Mag.}, vol.~56, no.~9,
  pp. 162--169, Sept. 2018.

\bibitem{survey2022mmWave}
Z.~Xiao, L.~Zhu, Y.~Liu, P.~Yi, R.~Zhang, X.-G. Xia, and R.~Schober, ``A survey
  on millimeter-wave beamforming enabled uav communications and networking,''
  \emph{IEEE Communications Surveys \& Tutorials}, vol.~24, no.~1, pp.
  557--610, 2022.

\bibitem{IRS_channel_model}
W.~Tang, M.~Z. Chen, X.~Chen, J.~Y. Dai, Y.~Han, M.~Di~Renzo, Y.~Zeng, S.~Jin,
  Q.~Cheng, and T.~J. Cui, ``Wireless communications with reconfigurable
  intelligent surface: Path loss modeling and experimental measurement,''
  \emph{IEEE Trans. Wireless Commun.}, vol.~20, no.~1, pp. 421--439, Jan. 2021.

\bibitem{rappaBook}
T.~S. Rappaport, \emph{Wireless Communications: Principles and Practice}.\hskip
  1em plus 0.5em minus 0.4em\relax New Jersey: Prentice Hall, 1996, 2nd Ed.

\bibitem{HoangTuy:Globaloptimization}
H.~Tuy, \emph{Convex Analysis and Global Optimization}.\hskip 1em plus 0.5em
  minus 0.4em\relax Springer Publishing Company, Inc., 2016.

\bibitem{DinhJSAC18}
V.-D. Nguyen, H.~V. Nguyen, O.~A. Dobre, and O.-S. Shin, ``A new design
  paradigm for secure full-duplex multiuser systems,'' \emph{IEEE J. Select.
  Areas Commun.}, vol.~36, no.~7, pp. 1480--1498, July 2018.

\bibitem{sedumi}
Y.~Labit, D.~Peaucelle, and D.~Henrion, ``{SEDUMI INTERFACE} 1.02: A tool for
  solving {LMI} problems with {SEDUMI},'' pp. 272--277, Oct. 2002.

\bibitem{WCNC_IRS_GreenNetwork}
S.~Sun, M.~Fu, Y.~Shi, and Y.~Zhou, ``Towards reconfigurable intelligent
  surfaces powered green wireless networks,'' in \emph{Proc. 2020 IEEE Wireless
  Commun. Net. Conf. (WCNC)}, Seoul, Korea, May 2020, pp. 1--6.

\end{thebibliography}
